\documentclass{article}
\usepackage{abstract}
\newcommand{\keywords}[1]{\par\noindent \textbf{Keywords:} #1}
\usepackage[english]{babel}
\usepackage{booktabs}
\usepackage[letterpaper,top=2cm,bottom=2cm,left=3cm,right=3cm,marginparwidth=1.75cm]{geometry}

\usepackage{amsmath}   
\usepackage{amsthm}
\newtheorem{theorem}{Theorem}

\usepackage{amsmath,amssymb,amsthm}
\usepackage{amsfonts}
\usepackage{bm}
\usepackage{multirow}
\usepackage{titling}
\usepackage{algorithm}
\usepackage{algorithmicx}  
\usepackage{algpseudocode}
\usepackage{comment}

\usepackage{graphicx}
\usepackage[colorlinks=true, allcolors=blue]{hyperref}
\usepackage{subcaption} 
\usepackage{float}  
\usepackage{float}           

\title{Physics-Informed Chebyshev Polynomial Neural Operator for Parametric Partial Differential Equations}
\author{Biao Chen, Jing Wang*, Hairun Xie, Qineng Wang, Shuai Zhang, Yifan Xia, Jifa Zhang*}

\begin{document}
\maketitle
\begin{abstract}
Neural operators have emerged as powerful deep learning frameworks for approximating solution operators of parameterized partial differential equations (PDE). However, current methods predominantly rely on multilayer perceptrons (MLPs) for mapping inputs to solutions, which impairs training robustness in physics-informed settings due to inherent spectral biases and fixed activation functions. 
To overcome the architectural limitations, we introduce the Physics-Informed Chebyshev Polynomial Neural Operator (CPNO), a novel mesh-free framework that leverages a basis transformation to replace unstable monomial expansions with the numerically stable Chebyshev spectral basis. By integrating parameter dependent modulation mechanism to main net, CPNO constructs PDE solutions in a near-optimal functional space, decoupling the model from MLP-specific constraints and enhancing multi-scale representation. 
Theoretical analysis demonstrates the Chebyshev basis's near-minimax uniform approximation properties and superior conditioning, with Lebesgue constants growing logarithmically with degree, thereby mitigating spectral bias and ensuring stable gradient flow during optimization. Numerical experiments on benchmark parameterized PDEs show that CPNO achieves superior accuracy, faster convergence, and enhanced robustness to hyperparameters. The experiment of transonic airfoil flow has demonstrated the capability of CPNO in characterizing complex geometric problems.

\end{abstract}


\section{Introduction}

Partial differential equations (PDEs) are fundamental mathematical tools for describing complex phenomena across a multitude of scientific and engineering disciplines~\cite{2022AGUFM.H52N0633D,DELCEY2024117389,FAROUGHI2020104218,math12010063,YAN2024104731,YAN2024123179}, and solving PDEs with neural networks has been widely used in fluid dynamics~\cite{StröferCiCP-25}, solid mechanics~\cite{HU2024112495}, materials science~\cite{zhang_dynamic_2024}, earth science~\cite{XU2021103828}, biomedical engineering~\cite{sel_physics-informed_2023}, and financial mathematics~\cite{GATTA202368}. Parametric PDEs capture how system behavior responds to variations in physical parameters, geometric configurations, boundary conditions, or initial states, which can be particularly crucial in applications such as climate modeling~\cite{Hourdin2020ProcessBasedCM} and aerospace engineering~\cite{Han_Hu_Jiang_Lee_2024}. However, solutions to these equations often exhibit multiscale characteristics and intricate patterns, such as spatially localized phenomena and parameter-dependent bifurcations. Consequently, the efficient and accurate solution of PDEs is paramount for understanding, predicting, and optimizing these intricate systems~\cite{WANG2021114037,SONG2023129493}.

\begin{figure}[htbp]
    \centering    
     \includegraphics[width=0.7\textwidth]{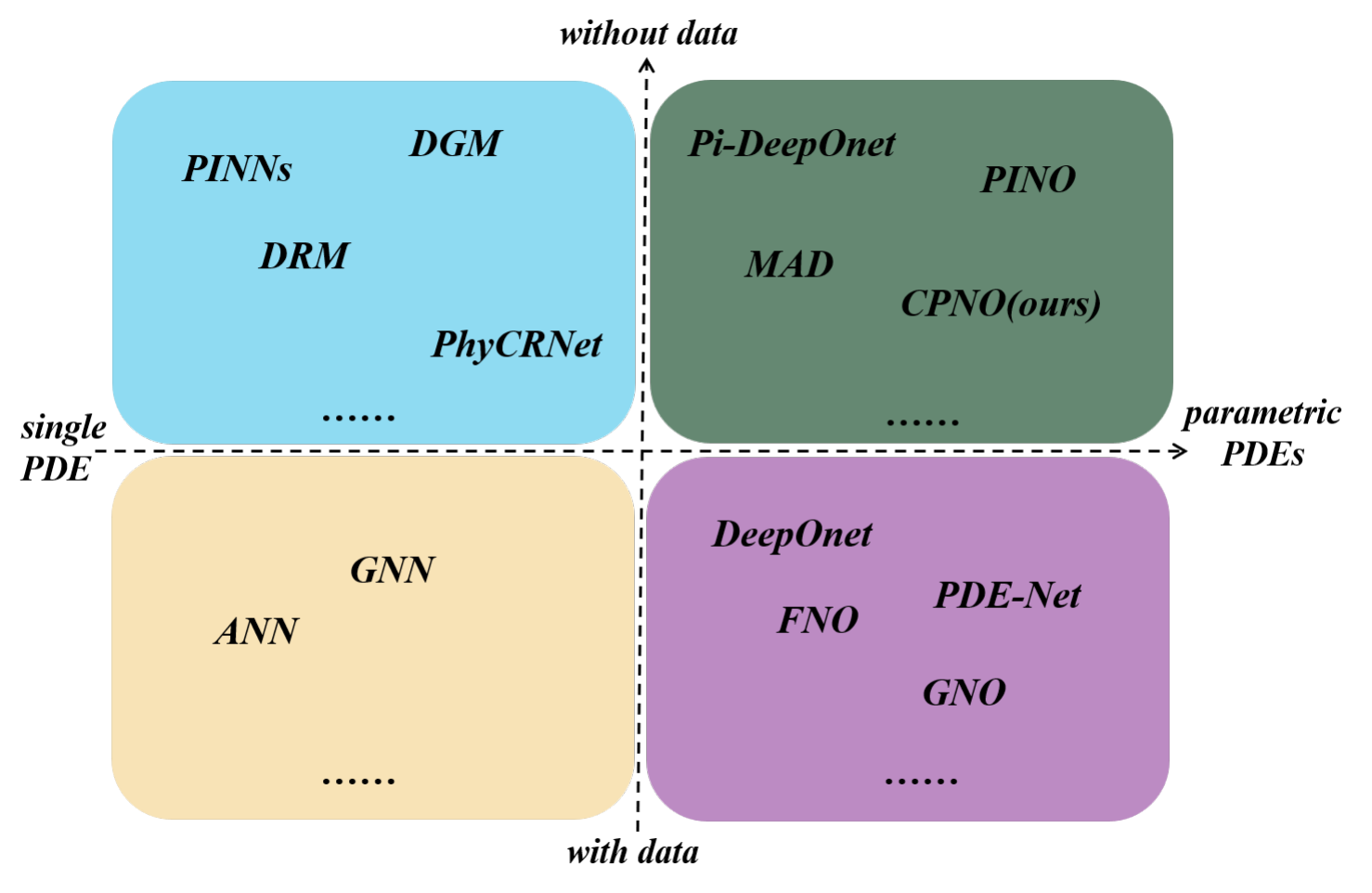} \hfill
    \caption{Classification of PDE solving methods by their ability to handle single vs parametric PDEs and data requirements.}
    \label{fig:pde-compare}
\end{figure}

Progress in deep learning has given rise to a spectrum of neural-network-based strategies for solving PDEs. These approaches for PDE solving can be naturally organized along two fundamental axes: whether the training relies on extensive datasets of pre-computed solutions, and whether the goal is to solve a single PDE or a family of parameterized PDEs~\cite{DBLP:journals/corr/abs-1910-03193, DBLP:journals/corr/abs-2003-03485, bhattacharya2021modelreductionneuralnetworks,lu2021learning}. Methods that require large supervisory datasets include classical Physics-Informed Neural Networks for individual problems as well as data-driven neural operators such as DeepONet~\cite{LU2024105282} and the Fourier Neural Operator (FNO)~\cite{li2020fourier} for parametric families. In contrast, purely physics-informed approaches dispense with solution data altogether, embedding the governing equations, boundary, and initial conditions directly into the loss function~\cite{RAISSI2019686}, this branch encompasses a rich variety of enhanced PINNs for single instances~\cite{WU2023115671,YU2022114823}. Beside, physics-informed neural operators, such as PINO~\cite{DBLP-pino}, physics-informed DeepONet~\cite{wang_learning_2021}, meta-auto encoder~\cite{MADHuang2021}, are capable of learning solution operators without any labeled solutions~\cite{DBLP-hyperpinn}. Figure~\ref{fig:pde-compare} illustrates this landscape as a quadrant diagram, with representative methods placed accordingly.
The present work is positioned in the most challenging—yet scientifically most rewarding quadrant: learning parametric solution operators with physics loss alone. 

In practice, physics-informed operator learning for parametric PDEs remains notoriously challenging. Recent studies of the training dynamics have shown that, under purely physics-based supervision, the loss landscape typically exhibits sharp, high-curvature ravines coupled with persistent high-frequency noise in the PDE residual~\cite{wang2020understanding,krishnapriyan2021,NIU2025130167}. This pathological behavior stems fundamentally from the widespread reliance on MLPs with fixed pointwise nonlinearities. Such architectures suffer from a pronounced spectral bias that preferentially captures low-frequency components, while struggling to represent the sharp gradients and multi-scale features that are characteristic of many PDE solutions~\cite{rahaman2019spectral,wang2020fail}.
A large body of work has sought to mitigate these difficulties through remedial interventions at the level of loss design or optimization. Prominent examples include causally weighted and Sobolev-norm losses~\cite{wang2024respect,YU2022114823}, residual-based adaptive refinement and importance sampling~\cite{wong2025evolution, CHANG2024325}, sequence-to-sequence training schedules~\cite{HONG2025100774}, and adversarial or variational formulations~\cite{shi2025whitebox}. Although these techniques can deliver noticeable gains on particular benchmarks, they remain essentially post-hoc corrections that leave the underlying architectural limitation untouched: the function space spanned by conventional MLPs is inherently prone to severe gradient imbalance and optimization stiffness.
An alternative and more direct strategy is to abandon pointwise MLPs entirely and parameterize the solution operator explicitly in a polynomial basis~\cite{singh2023poly}. While this approach circumvents the spectral bias of MLPs, naive implementations that employ the monomial basis reintroduce the well-known numerical ill-conditioning of high-degree polynomial interpolation, characterized by Gram matrices whose condition numbers grow exponentially with degree and by extreme sensitivity to coefficient perturbations~\cite{guo2025kolmogorovar,Yu_2026, yin2024chebyshev, jianan2023case,Brubeck2021}.

To achieve intrinsically stable and high-accuracy physics-informed operator learning, an architectural prior is needed that preserves the universal approximation capability of polynomials while imposing strict control over numerical conditioning from the very beginning—entirely sidestepping the need for intricate loss reweighting or adaptive sampling strategies. Approximation theory provides a definitive resolution to this challenge in the form of the Chebyshev polynomials of the first kind. On the reference interval  \([-1, 1]\), this basis is the unique minimax polynomial family: for any fixed degree p, it delivers the smallest possible maximum deviation from the best approximation among all polynomials of that degree~\cite{EAJAM-13-2,HAJIMOHAMMADI2021111530,yin2024chebyshevspectralneuralnetworks,SIVALINGAM2024150}. Crucially, its cosine-distributed roots cluster nodes near the domain boundaries, effectively eliminating the Runge phenomenon that plagues equidistance or monomial-based interpolation. As a direct consequence, the condition number of the associated discrete systems scales merely as $O(p^2)$ with polynomial degree p, a remarkably mild growth compared to the exponential $O(e^p)$ explosion exhibited by the monomial basis~\cite{Trefethen2013,Dardery2014,Mason2002,Nyengeri2021}.
By constructing the neural operator directly upon the Chebyshev spectral basis and complementing it with a lightweight, parameter-dependent modulation mechanism, the proposed Chebyshev Polynomial Neural Operator (CPNO) operates a prior within a provably well-conditioned and spectrally balanced function space. This foundational architectural decision removes the principal sources of training pathology at their root, yielding robust convergence under purely physics-based supervision and graceful incorporation of sparse observational data whenever available as rigorously demonstrated in Sections 2.5.

Building upon these insights, we propose a novel Physics-Informed Chebyshev polynomials Neural Operator (CPNO). The main contributions of this work are summarized as follows:

\begin{itemize}
    \item The implementation of a robust and efficient Chebyshev polynomial neural operator framework has been accomplished with notable success. The framework systematically constructs the representation of solutions in a function basis that is nearly optimal and numerically stable. This provides a new design paradigm that transcends the limitations of traditional fixed activation functions.
    \item We provide a rigorous theoretical analysis of the uniform approximation properties and numerical stability of the Chebyshev basis. The analysis elucidates how these properties directly address and resolve the prevalent issues of spectral bias and training instability found in existing neural operators.
    \item We demonstrate through extensive numerical experiments on several benchmark parameterized PDEs that CPNO significantly outperforms current physics-informed, mesh-free neural operator methods in terms of solution accuracy, convergence speed, and robustness.
\end{itemize}

As a physics-informed, mesh-free methodology, CPNO is designed to offer an efficient, accurate, and robust solution pathway for challenging parameterized systems. The rest of this paper is organized as follows: Section 2 introduce operator learning for parametric PDEs and give the details of model architecture. Section 3 introduce the theoretical advantages of Chebyshev polynomials. Section 4 presents and analyzes the numerical experiment results, including comparisons with baseline methods and ablation studies. Section 5 concludes the paper and discusses future research directions.

\section{Methodology}
This section details the Physics-Informed Chebyshev Polynomial Neural Operator (CPNO), a mesh-free architecture that employs Chebyshev spectral features to approximate solutions for parameterized PDEs. The core components of the model architecture and the physics-informed training paradigm are described.

\subsection{Problem Definition}
This subsection defines the general class of parameterized PDEs addressed by the framework and formalizes the solution operator to be approximated.
A general class of parameterized partial differential equations (PDEs) is considered, defined over a temporal domain $[0, T]$ and a spatial domain $\Omega \subset \mathbb{R}^d$ . The objective is to identify a solution $u: [0, T] \times \Omega \to \mathbb{R}$ residing in a suitable Banach space $V$. The system is governed by parameters $\theta$ drawn from a parameter space $\Theta$. For a given $\theta \in \Theta$, the solution $u(\cdot; \theta) \in V$ satisfies:
\begin{equation}
\begin{cases}
\mathcal{L}(\theta) u(x; \theta) = f(x; \theta), & x \in (0, T] \times \Omega, \\
\mathcal{B}(\theta) u(x; \theta) = g(x; \theta), & x \in (0, T] \times \partial \Omega, \\
\mathcal{I}(\theta) u(x; \theta) = u_0(x; \theta), & x \in \{0\} \times \Omega,
\end{cases}
\end{equation}

where $\mathcal{L}(\theta)$, $\mathcal{B}(\theta)$, and $\mathcal{I}(\theta)$ denote the differential, boundary, and initial operators, respectively, which may be nonlinear and parameter-dependent.
The parameter space $\Theta$ is formulated as a product space to unify sources of variation: $\theta = (a, f, g, u_0)$, where $a \in \mathcal{A}$ parameterizes the coefficients of $\mathcal{L}$ with $\mathcal{A}$ a function space on $\Omega$, $f \in \mathcal{F}$ ($L^2((0, T) \times \Omega)$ for the source term, $g \in \mathcal{G}$ ($L^2((0, T) \times \partial \Omega)$) for Dirichlet boundaries), and $u_0 \in \mathcal{U}_0$ (a function space on $\Omega$). The space $\Theta = \mathcal{A} \times \mathcal{F} \times \mathcal{G} \times \mathcal{U}_0$ is assumed compact and metric to ensure solution map stability and continuity.
The core task of operator learning is to approximate the solution operator $\mathcal{S}$ that maps each parameter instance $\theta \in \Theta$ to its unique weak solution $u(\cdot; \theta) \in V$:

\begin{equation}
\mathcal{S}: \Theta \to V, \quad \theta \mapsto \mathcal{S}(\theta) = u(\cdot; \theta).
\end{equation}
Under the well-posedness of the PDE, $\mathcal{S}$ is continuous and well-defined. A neural operator constructs a parameterized approximation $\mathcal{S}_\phi \approx \mathcal{S}$, with $\phi$ denoting the trainable network parameters.

\begin{figure}[htbp]
    \centering    
     \includegraphics[width=0.95\textwidth]{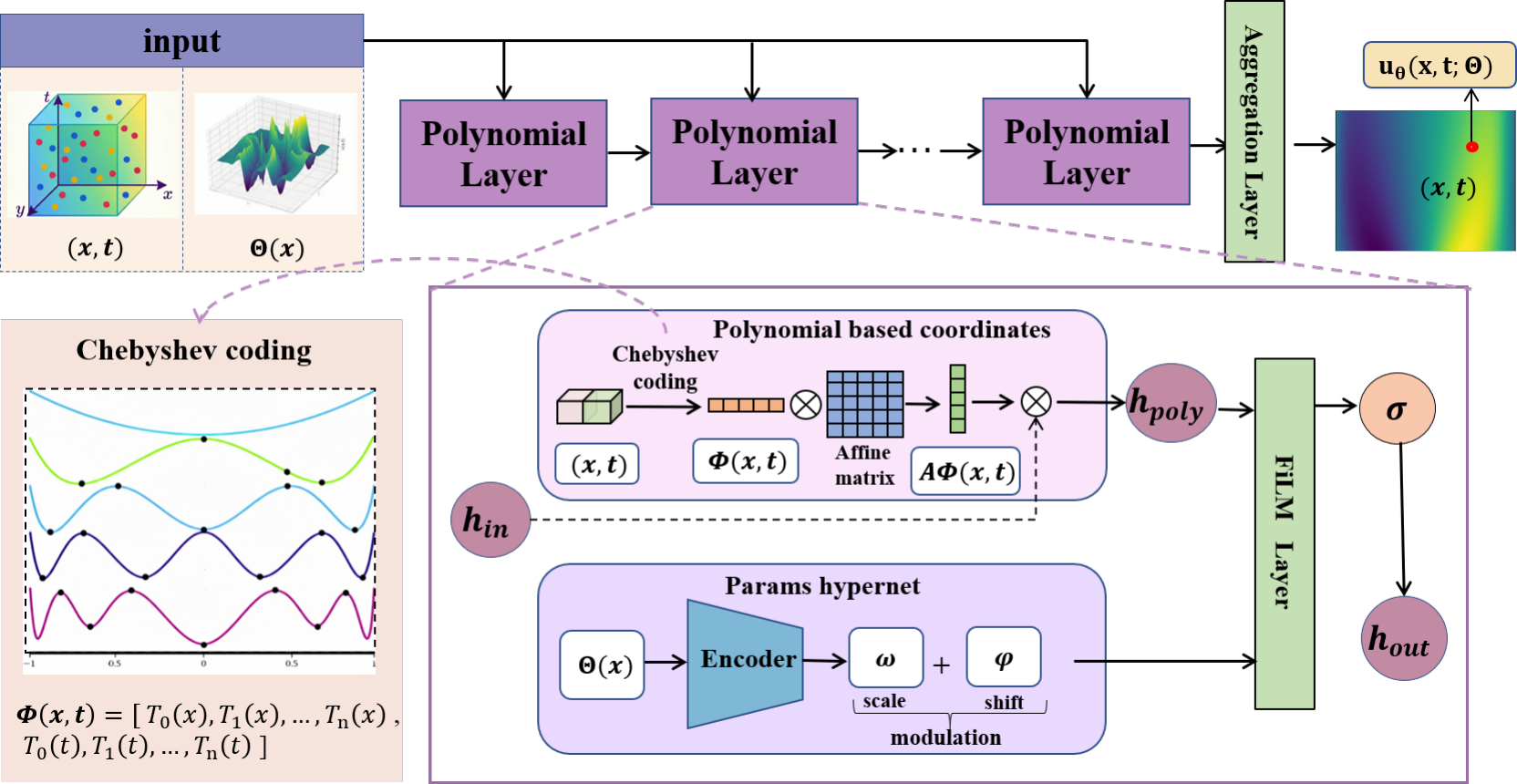} \hfill
    \caption{The main framework of the physics informed Chebyshev polynomial neural operator (CPNO)}
    \label{fig:main-frame}
\end{figure}

\subsection{Polynomial Stacking Neural Operator Framework}
We introduce a multi-layer stacking framework based on polynomial neural operators for the efficient solution of parametric PDEs. The architecture of the proposed CPNO is schematically depicted in Figure~\ref{fig:main-frame}. The framework takes spatial-temporal coordinates \((x, t) \in \Omega \times [0, T]\) and parameters \(\theta \in \Theta\) as inputs, iteratively constructing the solution \(u(x, t; \theta)\) through layer-wise transformations. Grounded in polynomial approximation theory, this approach leverages the nonlinear mapping of high-order polynomial spaces, inheriting the denseness property of the Weierstrass theorem, and is well-suited for capturing the multi-scale features of PDEs. Its modular design enables flexible adaptation of polynomial forms, theoretically approximating any continuous solution with robust generalization.

The proposed framework is a multi-layer neural operator architecture comprising \(q\) transformation mapping \(\{\mathcal{F}_j\}_{j=1}^q\), designed to approximate solutions to parameterized partial differential equations (PDEs). The process initiates with \(\mathbf{h}^{(0)} = \Phi(x, t)\), where \(\Phi(x, t)\) serves as the initial hidden state, encoding spatiotemporal coordinates \((x, t)\) through Chebyshev spectral features. The layer-wise evolution is governed by the recursive relation:
\begin{equation}
\mathbf{h}^{(j)} = \mathcal{F}_j(\mathbf{h}^{(j-1)}; \theta), \quad j = 1, 2, \dots, q,
\end{equation}
where \(\theta\) denotes the global PDE parameters influencing the operator’s behavior. The final solution is obtained via a linear projection:
\begin{equation}
u(x, t; \theta) = \mathbf{W}_q \mathbf{h}^{(q)} + b_q,
\end{equation}
with \(\mathbf{W}_q \in \mathbb{R}^{d_u \times d_h}\) and \(b_q \in \mathbb{R}^{d_u}\) representing the weight matrix and bias vector, respectively, mapping the terminal hidden state \(\mathbf{h}^{(q)} \in \mathbb{R}^{d_h}\) to the output space \(\mathbb{R}^{d_u}\).

Each transformation mapping \(\mathcal{F}_j\) is defined as a parameter-dependent nonlinear mapping:
\begin{equation}
\mathcal{F}_j(\mathbf{h}^{(j-1)}; \theta) = \sigma \left( \mathbf{f}_m^\theta \left( \mathbf{h}^{(j-1)} \right) \right),
\end{equation}
where \(\mathbf{f}_m^\theta\) denotes a modulation function dependent on \(\theta\), and \(\sigma\) is a nonlinear activation function, facilitating the adaptive adjustment of the hidden state across layers. The sequential application of these operators yields a composite transformation:
\begin{equation}
\mathbf{h}^{(q)} = \mathcal{F}_q \circ \mathcal{F}_{q-1} \circ \cdots \circ \mathcal{F}_1 (\Phi(x, t); \theta),
\end{equation}
which, when combined with the linear projection, produces the solution:
\begin{equation}
u(x, t; \theta) = \mathbf{W}_q \left( \mathcal{F}_q \circ \mathcal{F}_{q-1} \circ \cdots \circ \mathcal{F}_1 (\Phi(x, t); \theta) \right) + b_q.
\end{equation}
This layered composition effectively constructs a polynomial structure, with the explicit representation unrolling the recursion:
\begin{equation}
u(x, t; \theta) = \mathbf{W}_q \left( \sigma \left( \mathbf{f}_m^\theta \left[ \sigma \left( \cdots \sigma \left( \mathbf{f}_m^\theta \left( \Phi(x, t) \right] \right) \cdots \right) \right) \right) \right) ,
\end{equation}
where the nested application of \(\sigma \circ \mathbf{f}_m^\theta\) occurs \(q\) times, starting from the Chebyshev-encoded input \(\Phi(x, t)\). The resulting polynomial expansion leverages the denseness of Chebyshev polynomials on \([-1, 1]\), as per the Weierstrass approximation theorem, enabling the framework to approximate a broad class of PDE solutions with controllable accuracy through the depth \(q\) and parameter optimization.

\subsection{Polynomial Layer: Chebyshev Basis and Modulated Transformation}

Within the proposed framework, the \(j\)-th layer transformation operator \(\mathcal{F}_j\) serves as a pivotal component of the overall multi-layer stacking structure, mapping the previous layer output \(\mathbf{h}^{(j-1)}\) to \(\mathbf{h}^{(j)}\) according to \(\mathbf{h}^{(j)} = \mathcal{F}_j(\mathbf{h}^{(j-1)}; \theta)\). This layer leverages a synergistic design of Chebyshev spectral encoding and modulated transformation to capture the multi-scale features of parameterized partial differential equation (PDE) solutions, enhancing learning efficiency and numerical stability. 

\paragraph{Chebyshev Spectral Basis}
This sub-layer transforms low-dimensional spatiotemporal coordinates $(x, t)$ with $x = (x_1, \dots, x_d) \in \Omega$ into a high-dimensional spectral feature representation to resolve multi-scale characteristics. Each spatial coordinate component and the temporal coordinate undergo an affine transformation to normalize to the canonical domain of Chebyshev polynomials, $[-1, 1]$:

\begin{equation}
    x_i' = 2 \frac{x_i - x_{i,\min}}{x_{i,\max} - x_{i,\min}} - 1, \quad t' = 2 \frac{t - t_{\min}}{t_{\max} - t_{\min}} - 1
\end{equation}

where $x_{i,\min}$ and $x_{i,\max}$ denote the bounds of $\Omega$ for dimension i, and $t_{\min}$ and $t_{\max}$ denote the bounds of $[0, T]$. The encoding employs the three-term recurrence relation of Chebyshev polynomials of the first kind $T_k(z)$:

\begin{equation}
    T_0(x) = 1, \quad T_1(x) = x, \quad T_{k+1}(x) = 2x T_k(x) - T_{k-1}(x), \quad k \geq 1,
\end{equation}

computing values up to degree P for each normalized coordinate $x_i'$ and $t'$. The resulting sequences are concatenated across all components to form a spectral feature vector $\Phi(x, t)$:

\begin{equation}
  \Phi(x, t) = [T_0(x_1'), \dots, T_P(x_1'), \dots, T_0(x_d'), \dots, T_P(x_d'), T_0(t'), \dots, T_P(t')]^T \in \mathbb{R}^{(d+1)(P+1)}.  
\end{equation}

The orthogonality of Chebyshev polynomials (\(\int_{-1}^{1} T_j(x) T_k(x) (1 - x^2)^{-1/2} dx = \delta_{jk} / \pi\)) ensures numerical stability, while the recursive computation alleviates the network's burden of learning basis functions from raw coordinates, as noted in prior work. The feature vector \(\Phi(\mathbf{x}, t)\) serves as the input to the subsequent modulation sub-layer, significantly enhancing multi-scale representation.

\paragraph{Modulated Transformation} 

This sub-layer systematically constructs high-order polynomial terms by recursively combining the hidden state from the previous layer, $\mathbf{h}^{(l-1)}$, with the input Chebyshev spectral features $\Phi(\mathbf{x},t)$. First, a learnable linear transformation $\mathbf{A}_l$ maps $\Phi(\mathbf{x},t)$ to a space with the same dimension as the hidden state. The result is then combined with the previous hidden state via Hadamard product to increase the polynomial degree:

\begin{equation}
    \mathbf{h}^{(j)}(\mathbf{x},t) = (\mathbf{A} \Phi(\mathbf{x},t)) \odot \mathbf{h}^{(j-1)}(\mathbf{x},t),
\end{equation}
where $\mathbf{h}^{(0)}$ is typically initialized to a constant vector. This recursively generated feature, $\mathbf{h}^{(j)}$, which is rich in high-order cross-terms, is then dynamically modulated by the parameter $\theta$.

The modulation is performed using an amplitude vector $\boldsymbol{\omega}_l(\theta) \in \mathbb{R}^{d_h}$ and a phase vector $\boldsymbol{\varphi}_l(\theta) \in \mathbb{R}^{d_h}$, which are generated by the parameter-mapping network, followed by a nonlinear activation $\sigma$:

\begin{equation}
    \mathbf{h}^{(j)}(\mathbf{x},t; \theta) = \sigma \left( \boldsymbol{\omega}_l(\theta) \odot \mathbf{h}^{(j)}(\mathbf{x},t) + \boldsymbol{\varphi}_j(\theta) \right)
\end{equation}

Here, we employ the Gaussian Error Linear Unit (GELU) as the activation function $\sigma$. This design decouples the augmentation of polynomial degree from the parametric modulation, allowing the network to adaptively tailor its representation for specific PDE solutions within a structured and progressively enriched feature space. The design supports flexibility, where the $\Phi(\mathbf{x},t)$ component can be replaced by alternative spectral encodings like Legendre basis, while the modulation with $\boldsymbol{\omega}_l(\theta)$ and $\boldsymbol{\varphi}_l(\theta)$ preserves its core parameter-adaptive property. 

Based on the discussion above, we summarize the forward algorithm in algorithm\ref{alg:cpno}.

\begin{algorithm}[t]
\caption{Forward and Backward Pass for CPNO}
\label{alg:cpno}
\begin{algorithmic}[1]
\Require Spatiotemporal coordinates $\mathbf{x} \in \mathbb{R}^d$, PDE parameters $\theta$, Mapping network $\phi$, Synthesis network parameters $\psi = \{A_j, \omega_j, \phi_j\}_{j=1}^q, W_q$, Physics-informed loss $\mathcal{L}$
\Ensure Predicted solution $u(\mathbf{x}; \theta)$, Updated parameters $\phi, \psi$
\Statex \textbf{Forward Pass:}
\State $x' \gets$ Normalize($x$), $t' \gets$ Normalize($t$)  \Comment{Map to $[-1, 1]^d$ and $[-1, 1]$}
\State $\Phi(x, t) \gets \bigoplus_{i=1}^d [T_0(x_i'), \dots, T_P(x_i')] \oplus [T_0(t'), \dots, T_P(t')]$  \Comment{Chebyshev features}
\State $\mathbf{h}^{(0)} \gets \Phi(x, t) \in \mathbb{R}^{d_h}$  \Comment{Initialize hidden state}
\State for j = 1 to q do
\State $\mathbf{h}^{(j)}_{\text{poly}} \gets (A_j \Phi(x, t)) \odot \mathbf{h}^{(j-1)}$  \Comment{Polynomial terms}
\State $m_j \gets M_{\varphi,j}(\theta)$  \Comment{Modulation vector}
\State $\omega_j \gets \omega_{,j}(m_j)$; $\phi_j \gets \phi_{,j}(m_j)$  \Comment{Scale and shift}
\State $\mathbf{h}^{(j)} \gets \sigma(\omega_j \odot \mathbf{h}^{(j)}_{\text{poly}} + \phi_j)$  \Comment{Modulated activation}
\State end for
\State $u(x, t; \theta) \gets W_q \mathbf{h}^{(q)}$   \Comment{Project to solution}

\end{algorithmic}
\end{algorithm}

\subsection{Optimization Objectives and Loss Functions}
This subsection outlines the optimization objectives and loss functions used to train the CPNO parameters, ensuring compliance with physical constraints and incorporation of available data.
The trainable parameters $\phi$ of CPNO are optimized by minimizing a composite loss function $\mathcal{L}(\phi)$. This function enforces adherence to the physical laws of the PDE system and integrates observational data when available. The total loss is expressed as a weighted sum of four components:

\begin{equation}
\mathcal{L}(\phi) = \lambda_{\text{pde}} \mathcal{L}_{\text{pde}}(\phi) + \lambda_{\text{ic}} \mathcal{L}_{\text{ic}}(\phi) + \lambda_{\text{bc}} \mathcal{L}_{\text{bc}}(\phi) + \lambda_{\text{data}} \mathcal{L}_{\text{data}}(\phi),    
\end{equation}

where $\lambda_{\text{pde}}$, $\lambda_{\text{ic}}$, $\lambda_{\text{bc}}$, and $\lambda_{\text{data}}$ denote user-defined hyperparameters that balance each term's contribution.
The PDE residual loss $\mathcal{L}_{\text{pde}}$ ensures that the predicted solution $u_\phi(x, t; \theta)$ satisfies the governing differential equation. It is computed as the mean squared error of the PDE residual, evaluated over collocation points sampled from the spatiotemporal domain and parameter space. Differential operators in $\mathcal{L}(\theta)$ are obtained via automatic differentiation:

\begin{equation}
\mathcal{L}_{\text{pde}}(\phi) = \mathbb{E}_{(x,t) \sim \mathcal{U}(\Omega \times (0,T]), \theta \sim \mathcal{P}(\Theta)} \left[ \left\| \mathcal{L}(\theta) u_\phi(x, t; \theta) - f(x, t; \theta) \right\|_2^2 \right],
\end{equation}

where $\mathcal{U}$ represents uniform distribution over the domain and $\mathcal{P}(\Theta)$ denotes the distribution over the parameter space.
The initial condition loss $\mathcal{L}_{\text{ic}}$ and boundary condition loss $\mathcal{L}_{\text{bc}}$ enforce constraints on their respective domains. Each is defined as the mean squared error between predictions and prescribed values, promoting alignment with initial states and boundary behaviors:

\begin{equation}
\mathcal{L}_{\text{ic}}(\phi) = \mathbb{E}_{x \sim \mathcal{U}(\Omega), \theta \sim \mathcal{P}(\Theta)} \left[ \left\| \mathcal{I}(\theta) u_\phi(0, x; \theta) - u_0(x; \theta) \right\|_2^2 \right],
\end{equation}

\begin{equation}
\mathcal{L}_{\text{bc}}(\phi) = \mathbb{E}_{(x,t) \sim \mathcal{U}(\partial \Omega \times (0,T]), \theta \sim \mathcal{P}(\Theta)} \left[ \left\| \mathcal{B}(\theta) u_\phi(x, t; \theta) - g(x, t; \theta) \right\|_2^2 \right]
\end{equation}

When sparse solution measurements are available, the data fidelity loss $\mathcal{L}_{\text{data}}$ aligns predictions with ground-truth observations. This term minimizes discrepancies for known input-output pairs $(x, t, \theta, u_{\text{ref}})$ from dataset $\mathcal{D}$:

\begin{equation}
\mathcal{L}_{\text{data}}(\phi) = \mathbb{E}_{(x,t,\theta,u_{\text{ref}}) \sim \mathcal{D}} \left[ \left\| u_\phi(x, t; \theta) - u_{\text{ref}} \right\|_2^2 \right].
\end{equation}

The training objective identifies the optimal parameters $\phi^*$ that minimize the total loss:

\begin{equation}
\phi^* = \arg \min_\phi \mathcal{L}(\phi).
\end{equation}

This optimization is performed using stochastic gradient-based methods. The Adam optimizer is applied for its adaptive learning rates and effective handling of deep network training. At each step, the loss and gradients with respect to $\phi$ are calculated on mini-batches comprising collocation points and parameter samples from the respective distributions. Parameters are updated iteratively per the Adam rule. A learning rate annealing schedule such as exponential decay supports fine-tuning during later training stages, promoting convergence to an accurate and physically consistent solution operator.

\subsection{Theoretical Analysis of Chebyshev Polynomials Neural Operators}

This section provides a rigorous theoretical analysis establishing that the CPNO framework simultaneously achieves near-optimal expressivity and trainability. From the perspective of classical approximation theory, the Chebyshev spectral encoding yields approximation errors that converge at a rate arbitrarily close to that of the best possible polynomial approximation. Complementarily, from the viewpoint of optimization dynamics and numerical linear algebra, the Chebyshev basis exhibits near-optimal conditioning with Lebesgue constants growing merely logarithmically with degree, thereby effectively suppressing spectral bias, ensuring numerical stability in both forward and backward propagation, and guaranteeing robust, well-conditioned gradient flow throughout training—properties that are provably unattainable with the notoriously ill-conditioned monomial bases. These analytically derived attributes of spectral convergence and dynamical stability collectively underpin the framework’s superior accuracy, generalization performance, and training reliability observed empirically.

\subsubsection{Chebyshev Spectral Approximation and Error Decomposition}

To rigorously quantify the learning objective of the CPNO framework, we analyze the approximation error from the perspective of spectral projection. Consider the target solution operator $\mathcal{S}: \Theta \to V$ mapping a parameter instance $\theta$ to the solution $u(x;\theta)$. The neural operator constructs an approximation $u_{\phi}(x;\theta)$. Instead of assuming $u_{\phi}$ lies strictly within a polynomial space, we decompose the total error using the triangle inequality to isolate the contribution of the functional basis from the network's optimization capability:

\begin{equation}
\label{eq:error_decomposition}
\|u(x;\theta) - u_{\phi}(x;\theta)\|_{L^\infty} \le \|u(x;\theta) - \mathcal{P}_N u(x;\theta)\|_{L^\infty} + \|\mathcal{P}_N u(x;\theta) - u_{\phi}(x;\theta)\|_{L^\infty}
\end{equation}

where the first term is truncation error and second is estimation error. $\mathcal{P}_N$ denotes the optimal spectral projection operator onto the space of Chebyshev polynomials of degree up to $N$:

\begin{equation}
\mathcal{P}_N u(x;\theta) = \sum_{k=0}^N c_k^*(\theta) T_k(x), \quad c_k^*(\theta) = \frac{\langle u(\cdot;\theta), T_k \rangle_w}{\|T_k\|_w^2}
\end{equation}

Here, $\langle \cdot, \cdot \rangle_w$ represents the weighted inner product with weight $w(x) = (1-x^2)^{-1/2}$. The first term, $\mathcal{E}_{tru} = \|u(x;\theta) - \mathcal{P}_N u(x;\theta)\|_{L^\infty}$, represents the theoretical lower bound of the error, governed solely by the choice of basis and the regularity of the solution.

\begin{theorem}
[Exponential Decay of Truncation Error]
Let $u(x;\theta)$ be analytic within a Bernstein ellipse $\mathcal{E}_{\rho}$ with foci at $\pm 1$ and semi-major axis summed with semi-minor axis equal to $\rho > 1$. The truncation error of the Chebyshev series satisfies:
\begin{equation}
\mathcal{E}_{app} = \left\| u(x;\theta) - \sum_{k=0}^N c_k^*(\theta) T_k(x) \right\|_{L^\infty} \le \frac{C \rho^{-N}}{\rho - 1}
\end{equation}
where $C$ is a constant dependent on the maximum modulus of $u$ within $\mathcal{E}_{\rho}$.
\end{theorem}
Theorem 1 establishes the spectral convergence property of the CPNO backbone. The error decays exponentially with $N$, contrasting with the algebraic decay of grid-based methods. This ensures that a moderate $N$ is sufficient to capture complex solution manifolds. The Proof can be found in Appendix B.

\subsubsection{Estimation Error Bounds and Numerical Stability}

The second term of equation~\ref{eq:error_decomposition} is estimation error $\mathcal{E}_{est}$, quantifies the discrepancy between the optimal spectral projection and network output:
\begin{equation}
\mathcal{E}_{est} = \left\| \sum_{k=0}^N c_k^*(\theta) T_k(x) - u_{\phi}(x;\theta) \right\|_{L^\infty}
\end{equation}
Minimizing $\mathcal{E}_{est}$ requires the neural network to accurately approximate the mapping $\theta \mapsto \{c_k^*(\theta)\}_{k=0}^N$. This reduces the PDE solving task to a coefficient regression problem. A distinctive advantage of the CPNO framework lies in its inherent numerical stability, which can be rigorously controlled through bounds on the target function $u(x; \theta)$ and its derivatives.

\begin{theorem} [Bound on Function Amplitude]
By leveraging the property of Chebyshev polynomials, $|T_k(x)| \leq 1$ for all $x \in [-1, 1]$, the amplitude of target function satifies 

\begin{equation}
    |u_s(x; \boldsymbol{\theta})| = \left| \sum_{k=0}^{N} c_k(\boldsymbol{\theta}) T_k(x) \right| \leq \sum_{k=0}^{N} |c_k(\boldsymbol{\theta})| \cdot |T_k(x)| \leq \sum_{k=0}^{N} |c_k(\boldsymbol{\theta})|.
\end{equation}
Taking the $L^\infty$ norm yields a more compact bound:
\begin{equation}
    \|u_s(x; \boldsymbol{\theta})\|_{L^\infty([-1,1])} \leq \|\mathbf{c}(\boldsymbol{\theta})\|_1,
\end{equation}
where $\mathbf{c}(\theta)=\left[c_0(\theta), \ldots, c_N(\theta)\right]^T$ and $\|\cdot\|_1$ denotes the vector $\ell^1$ norm.. The global amplitude of $u_s$ is thereby governed by the $L^1$ norm of its spectral coefficient vector. 
\end{theorem}

By imposing constraints on the output of the network $\mathcal{M}_c$, such as a $GELU$ activation or norm clipping, the condition $\|\mathbf{c}(\boldsymbol{\theta})\|_1 \leq C$ can be enforced, thus guaranteeing the boundedness of $u_s$. This mechanism ensures the numerical stability of the forward pass and prevents solution divergence.

In a physics-informed setting, the stability of the solution's derivatives is of paramount importance. Defining the $m$-th order differential operator as $\mathcal{D}^m = \frac{d^m}{dx^m}$, we have:
\begin{equation}
    \mathcal{D}^m u_s(x; \boldsymbol{\theta}) = \sum_{k=0}^{N} c_k(\boldsymbol{\theta}) \mathcal{D}^m T_k(x).
\end{equation}

\begin{theorem} [Bound on Derivative Amplitude]
The $m$-th order derivative of a Chebyshev polynomial satisfies the bound $\|\mathcal{D}^m T_k(x)\|_{L^\infty} \leq C_m k^{2m}$, where $C_m$ is a constant dependent only on $m$. Consequently, the magnitude of the solution's derivative is bounded as:
\begin{equation}
    |\mathcal{D}^m u_s(x; \boldsymbol{\theta})| \leq \sum_{k=0}^{N} |c_k(\boldsymbol{\theta})| \cdot |\mathcal{D}^m T_k(x)| \leq C_m \sum_{k=0}^{N} |c_k(\boldsymbol{\theta})| k^{2m}.
\end{equation}
\end{theorem}

The magnitude of the solution's derivative is determined by a weighted $L^1$ norm of the spectral coefficients. To ensure the stability of high-order derivatives, the network is implicitly encouraged to learn spectrally decaying coefficients, where $|c_k(\boldsymbol{\theta})|$ decreases rapidly enough to counteract the polynomial growth of $k^{2m}$. This characteristic is consistent with the spectral decay behavior of solutions to real-world physical systems and robustly prevents numerical breakdown caused by high-order terms with large coefficients, thereby enhancing the numerical robustness of the forward propagation.

\subsubsection{Theoretical Analysis of Numerical Stability in Backward Propagation}

The advantages of our method are particularly pronounced in the numerical stability of the training process. We first consider the case of a linear differential operator $\mathcal{L}$ for a PDE defined as $\mathcal{L}[u](x) = f(x; \boldsymbol{\theta})$. The residual is given by:
\begin{equation}
    \mathcal{R}(x; \boldsymbol{\theta}) = \mathcal{L}[u_s](x; \boldsymbol{\theta}) - f(x; \boldsymbol{\theta}) = \sum_{k=0}^{N} c_k(\boldsymbol{\theta}) \mathcal{L}[T_k](x) - f(x; \boldsymbol{\theta}).
\end{equation}

The loss function is defined as the squared $L^2$ norm of the residual: $Loss = \int_{-1}^{1} (\mathcal{R}(x; \boldsymbol{\theta}))^2 dx$. The gradient with respect to a spectral coefficient $c_j(\boldsymbol{\theta})$ is:
\begin{equation}
    \frac{\partial Loss}{\partial c_j} = 2 \int_{-1}^{1} \mathcal{R}(x; \boldsymbol{\theta}) \frac{\partial \mathcal{R}}{\partial c_j} dx.
\end{equation}
Due to the linearity of the operator $\mathcal{L}$, we have $\frac{\partial \mathcal{R}}{\partial c_j} = \mathcal{L}[T_j](x)$, which is a fixed basis function. Letting $\phi_j(x) = \mathcal{L}[T_j](x)$, the gradient becomes:
\begin{equation}
    \frac{\partial Loss}{\partial c_j} = 2 \int_{-1}^{1} \left( \sum_{k=0}^{N} c_k(\boldsymbol{\theta}) \phi_k(x) - f(x; \boldsymbol{\theta}) \right) \phi_j(x) dx.
\end{equation}

The gradient vector $\nabla_{\mathbf{c}} Loss$ can be written in matrix form as:
\begin{equation}
    \nabla_{\mathbf{c}} Loss = 2 (\mathbf{G} \mathbf{c} - \mathbf{b}),
\end{equation}
where $\mathbf{G}$ is the Gram matrix with elements and $\mathbf{b}$ is a vector with elements 
$$
G_{jk} = \int_{-1}^{1} \phi_j(x) \phi_k(x) dx = \int_{-1}^{1} \mathcal{L}[T_j](x) \mathcal{L}[T_k](x) dx
$$
$$
b_j = \int_{-1}^{1} f(x) \phi_j(x) dx
$$
The core of the training stability lies in the spectral properties of the Gram matrix $\mathbf{G}$, particularly its condition number $\kappa(\mathbf{G})$. The properties of this matrix are determined by the choice of function basis and the linear operator $\mathcal{L}$. To fundamentally reveal the impact of the basis selection, we conduct a detailed comparison of the Gram matrices generated by the monomial and Chebyshev bases. The condition number, $\kappa(\mathbf{G})$, of the Gram matrix directly characterizes the geometry of the quadratic loss landscape. A fundamental analysis of the Gram matrices generated by the Chebyshev and the standard monomial bases can therefore reveal the intrinsic differences in their numerical stability. To elucidate these intrinsic properties, we consider the simplest yet most representative case: the identity operator, i.e., $\mathcal{L}=I$. In this scenario, the spectral expansion is directly employed to fit a target function.

We first analyze the standard monomial basis, $\{\psi_k(x) = x^k\}_{k=0}^N$. The elements of the corresponding Gram matrix, $\mathbf{G}_M$, are given by:
\begin{equation}
    G_{jk}^{(M)} = \langle x^j, x^k \rangle_{L^2} = \int_{-1}^{1} x^j x^k dx = \int_{-1}^{1} x^{j+k} dx.
\end{equation}
The integral can be computed directly:
\begin{equation}
    G_{jk}^{(M)} = \left[ \frac{x^{j+k+1}}{j+k+1} \right]_{-1}^{1} = \frac{1 - (-1)^{j+k+1}}{j+k+1},
\end{equation}
which can be further simplified to:
\begin{equation}
    G_{jk}^{(M)} =
    \begin{cases}
        \frac{2}{j+k+1} & \text{if } j+k \text{ is even}, \\
        0               & \text{if } j+k \text{ is odd}.
    \end{cases}
\end{equation}
This matrix possesses a structure and numerical properties remarkably similar to the infamous \textbf{Hilbert matrix} (whose elements are $H_{ij} = \frac{1}{i+j-1}$). Its condition number, $\kappa(\mathbf{G}_M)$, grows exponentially with the matrix size $N$, approximated by $\kappa(\mathbf{G}_M) \sim O(e^{3.5N})$. This indicates that the Gram matrix derived from the monomial basis is extremely ill-conditioned, implying that any learning framework based on high-order monomials is intrinsically numerically unstable.

Next, we analyze the Chebyshev basis, $\{\psi_k(x) = T_k(x)\}_{k=0}^N$. The elements of its Gram matrix, $\mathbf{G}_T$, are:
\begin{equation}
    G_{jk}^{(T)} = \langle T_j, T_k \rangle_{L^2} = \int_{-1}^{1} T_j(x) T_k(x) dx.
\end{equation}
It is crucial to note that this inner product is the standard, unweighted $L^2$ inner product, not the weighted inner product $\langle \cdot, \cdot \rangle_w$ (with weight $(1-x^2)^{-1/2}$) under which Chebyshev polynomials are perfectly orthogonal. Consequently, the resulting matrix is not strictly diagonal. The integral has a known analytical solution:
\begin{equation}
    G_{jk}^{(T)} =
    \begin{cases}
        0                               & \text{if } j+k \text{ is odd}, \\
        \frac{-2}{(j+k-1)(j+k+1)}       & \text{if } j \ne k \text{ and } j+k \text{ is even}, \\
        \pi                             & \text{if } j=k=0, \\
        \frac{\pi}{2}                   & \text{if } j=k>0.
    \end{cases}
\end{equation}
An analysis of this matrix structure reveals that the diagonal elements, $G_{kk}^{(T)}$, with values of $\pi$ or $\pi/2$, are the largest entries in the matrix. The off-diagonal elements, $G_{jk}^{(T)}$, decrease rapidly as a function of the distance from the diagonal, with a decay rate of $O((j+k)^{-2})$. The Gram matrix derived from the Chebyshev basis under the standard $L^2$ inner product is therefore diagonally dominant. Such matrices are known to be well-conditioned, and their condition number, $\kappa(\mathbf{G}_T)$, grows very slowly with $N$, typically at a low-order polynomial rate. This provides a solid mathematical foundation for the stability, convergence speed, and final accuracy of our operator framework. 

For a non-linear differential operator $\mathcal{N}$, although the loss function $L$ is no longer a simple quadratic form and the learning problem becomes a non-convex optimization, the Chebyshev basis continues to guarantee stability. First, the boundedness of the solution ($|T_k(x)| \leq 1$) and its derivatives ($|T_k'(x)| \leq k^2$) ensures that the computation of complex non-linear residuals does not fail due to the numerical explosion of the basis functions themselves. Second, its near-orthogonality ensures that the local sub-problems encountered during optimization are relatively well-conditioned, avoiding the pathological behavior that can arise from a monomial basis.

\section{Experiment Results and Discussions}
To rigorously assess our method, we conducted experiments across parametric PDEs, including Burgers Equation, Diffusion-Reaction System and Allen-Cahn equation. The parameter spaces for all problems were generated using a Gaussian Random Field (GRF) with a zero mean and an exponential quadratic kernel $k_l\left(x_1, x_2\right)=$ $\exp \left(-\frac{\left|x_1-x_2\right|^2}{2 t^2}\right)$. Our method is compared against physics-informed, mesh-free methods designed for querying solutions to parameterized PDEs, The baselines include PI-DeepONet, MAD without fine-tuning and Hyper-PINNs. For fair comparison, all methods were implemented using fully connected neural networks with 4 hidden layer and the width is 64. We utilized GELU activation functions and Adam optimizer.  All experiments were performed on a single NVIDIA V100 GPU. Performance is primarily evaluated using the relative $L_2$ error:
\begin{equation}
    \text{Error} = \frac{\|u_{\text{pred}} - u_{\text{true}}\|_{L_2}}{\|u_{\text{true}}\|_{L_2}} 
\end{equation}
where $u_{\text {true }}$ denotes the reference solution and $u_{\text {pred }}$ is the neural network prediction. This metric provides a normalized measure of solution accuracy, independent of the solution's magnitude, and is widely adopted in PDE-related tasks. Further details are available on \url{https://github.com/biao023/CPNO.git}.

Our experiments begin with a comparative analysis of CPNO's fundamental test results against baselines. Subsequently, we investigate the method's convergence characteristics, its efficacy in approximating multi-scale error flow fields. Finally, ablation study of the Chebyshev encoding order on solution accuracys are performed.

\subsection{Comparative Numerical Results}
To provide an initial assessment of the proposed CPNO for solving parameterized partial differential equations (PDEs), this section presents a direct comparison against several representative physics-informed, mesh-free methodologies prevalent in the field. These baseline methods include Physics-Informed DeepONet (PI-DeepONet), Meta-learning Assisted DeepONet (MAD), and Hypernetwork-based PINNs (HyperPINNs). For CPNO, we evaluate two distinct training paradigms: first, a purely physics-informed zero-shot learning approach (CPNO-zero), which does not rely on any observational data of the true solution; and second, a few-shot learning approach (CPNO-few), which augments the physics-informed training with a minimal set of observational data, specifically solutions corresponding to three parameter instances. Benchmark PDEs include burgers equation, Allen-Cahn equation, Diffusion-reaction system and Navier-Stokes equation, and their detailed description are provided in Appendix A.  

\begin{table}[h]
\centering
\caption{Relative $L2$ Errors of CPNO and baselines for Parametric PDEs}
\label{tab:basic-results}
\begin{tabular}{lccccc}
\toprule
Method & Burgers & Allen-Cahn & Diffusion-Reaction & Navier-Stokes\\
\midrule
PI-DeepoNet & 0.0809 & 0.0237 &  0.0318 & 0.0342\\
MAD         & 0.1063 & 0.2262 &  0.0750 & 0.1918\\
HyperPINNs  & 0.0455 & 0.0476 &  0.0314 & 0.0406\\
CPNO-zero & 0.0306 & 0.0169 & 0.0296 & 0.0272\\
CPNO-few & 0.0290 & 0.0109 & 0.0299 & 0.0166\\
\bottomrule
\end{tabular}
\end{table}

Table \ref{tab:basic-results} summarizes the relative L2 errors achieved by CPNO and the baseline methods across three benchmark problems, indicating that the proposed CPNO method demonstrates competitive performance across the tested parameterized PDE problems. In the purely physics-informed zero-shot setting (CPNO-zero), CPNO consistently surpasses all competing methods, yielding error reductions of up to 62\% relative to the strongest prior approach (HyperPINN) on the Navier–Stokes benchmark and establishing new best results across the entire suite. When merely a few ($\leq5\%$) sparsely sampled solution observations are incorporated during training, the few-shot variant (CPNO-few) further improves accuracy on three of the four problems, achieving the lowest errors reported to date on the Allen–Cahn(35\% improvement over CPNO-zero), diffusion–reaction, and Navier–Stokes (39\% improvement) equations. On the Burgers’ equation, CPNO-few remains within 5\% of its zero-shot counterpart while still substantially outperforming all baselines. These results underscore CPNO’s exceptional data efficiency, architectural expressiveness, and seamless ability to leverage limited supervisory signal atop a robust physics-informed foundation.
These initial findings suggest that the CPNO framework, whether operating in zero-shot mode or few-shot mode , can provide high-accuracy, mesh-free solutions for parameterized PDEs.

\subsection{Converge Speed and Accuracy Study}
To comprehensively evaluate the performance of our proposed CPNO and to rigorously validate the efficacy of its core component—the Chebyshev spectral encoding—we conduct a comparative analysis against a key ablation model. This baseline, denoted as w/o-CPNO, features an identical architecture but omits the spectral encoding module, processing raw spatiotemporal coordinates directly. This section provides a detailed comparison of the convergence rates and the training iterations required to achieve predefined accuracy thresholds for the CPNO methods and w/o-CPNO across all three benchmark parameterized PDE problems. All the training methods employed in this section are purely physics-driven.

\begin{figure}[htbp]
    \centering    
     \includegraphics[width=0.8\textwidth]{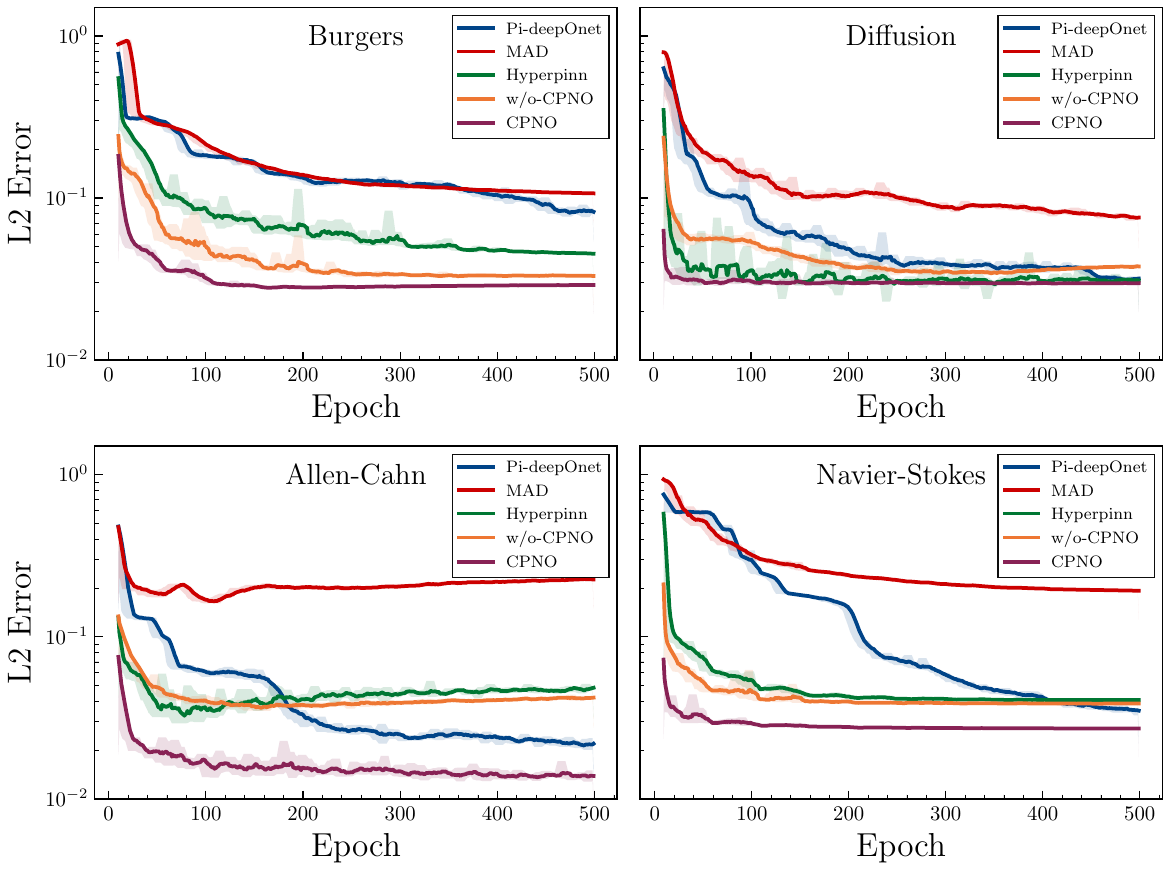} \hfill
    \caption{Comparison of Convergence Curves for Different Methods on test cases.}
    \label{fig:converge-all}
\end{figure}

Figure~\ref{fig:converge-all} illustrates the convergence curves for the different methodologies applied to each parameterized PDE. The abscissa represents the number of training epochs, while the ordinate denotes the corresponding relative $L_2$ test error. A consistent observation across all three benchmark problems is that the proposed CPNO exhibits the most rapid convergence and attains the highest final accuracy. Among the baseline methods, HyperPINN generally demonstrates the strongest performance, albeit with a significantly larger parameter count compared to other baselines. PI-DeepONet and MAD, in contrast, show comparatively slower convergence and achieve less favorable final accuracies. Notably, the w/o-CPNO, without Chebyshev coding enhancement, surpasses HyperPINN in performance on the Burgers' equation but is slightly outperformed by HyperPINN on the Allen-Cahn and diffusion-reaction systems. In contrast, the CPNO, benefiting from its enhanced input representation and efficient parameter modulation, consistently displays a rapid initial descent in its error curve, stabilizing at the lowest error levels within fewer training epochs. This underscores its comprehensive advantages in both learning efficiency and model expressivity.

\begin{figure}[htbp]
    \centering    
     \includegraphics[width=0.8\textwidth]{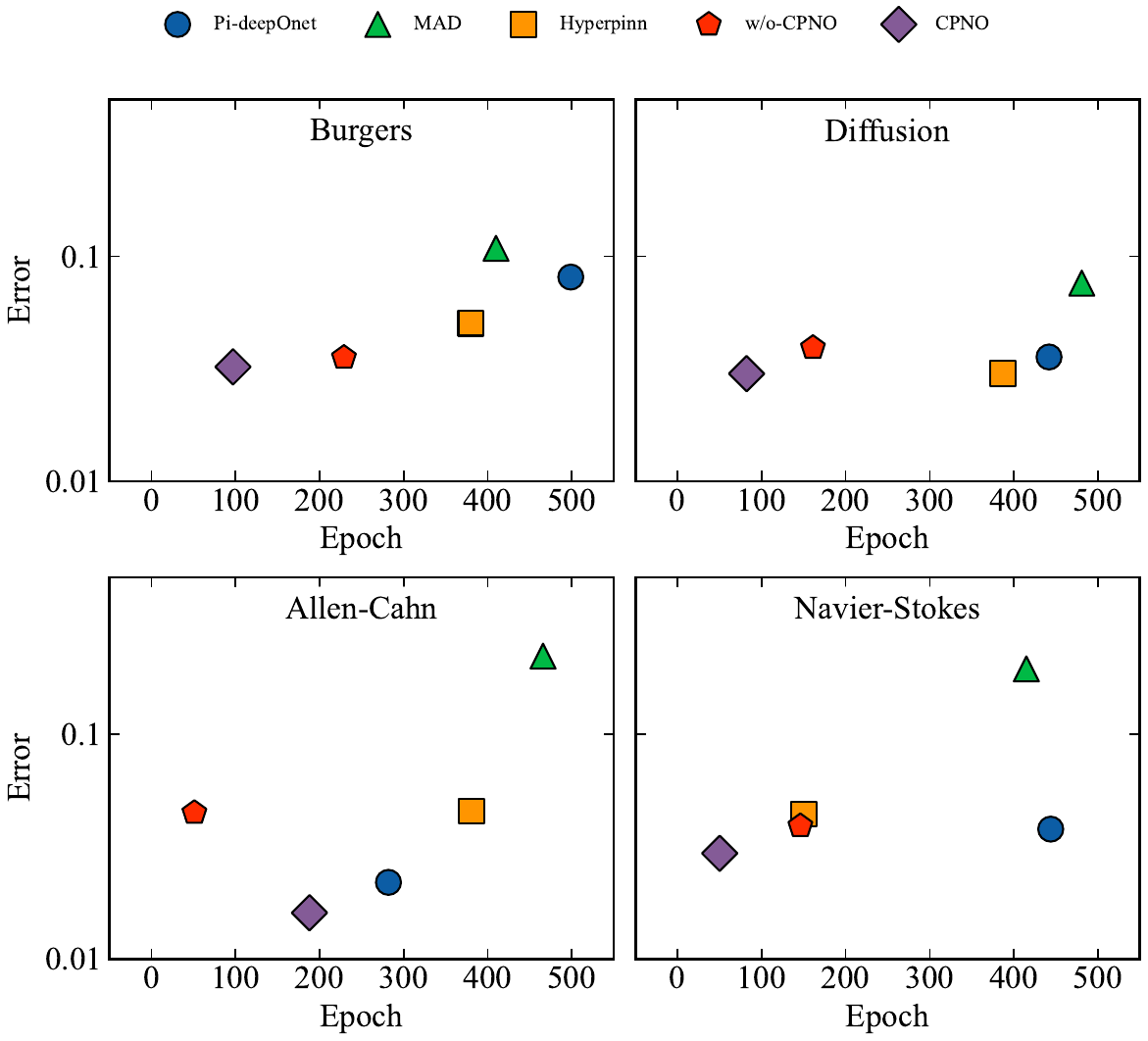} \hfill
    \caption{Comparison of Minimum Training Epochs Required to Reach Specific Test Accuracies for Different Methods on test cases.}
    \label{fig:efficiency-pareto}
\end{figure}

Figure~\ref{fig:efficiency-pareto} provides a holistic performance assessment by visualizing the trade-off between final solution accuracy and computational efficiency. In this plot, the converged relative $L_2$ test error for each method is plotted against its corresponding training cost, quantified by the number of epochs required for convergence. Optimal performance in this space corresponds to the region of low error and low computational cost, located at the bottom-left of the plot. Across all three benchmark problems, the proposed CPNO consistently occupies this optimal region, defining the Pareto front of the evaluated methodologies. While some baseline methods achieve comparable accuracy, they do so at a substantially higher training cost. Conversely, other methods that converge more rapidly are limited to a significantly higher final error. These results demonstrate that the CPNO framework achieves a superior balance of accuracy and efficiency, enabling the generation of high-fidelity solutions for parameterized PDEs with significantly reduced computational expenditure. This efficiency is particularly pertinent for practical applications requiring rapid model iteration or the exploration of extensive parameter spaces.

\begin{figure}[htbp]
    \centering    
     \includegraphics[width=1.0\textwidth]{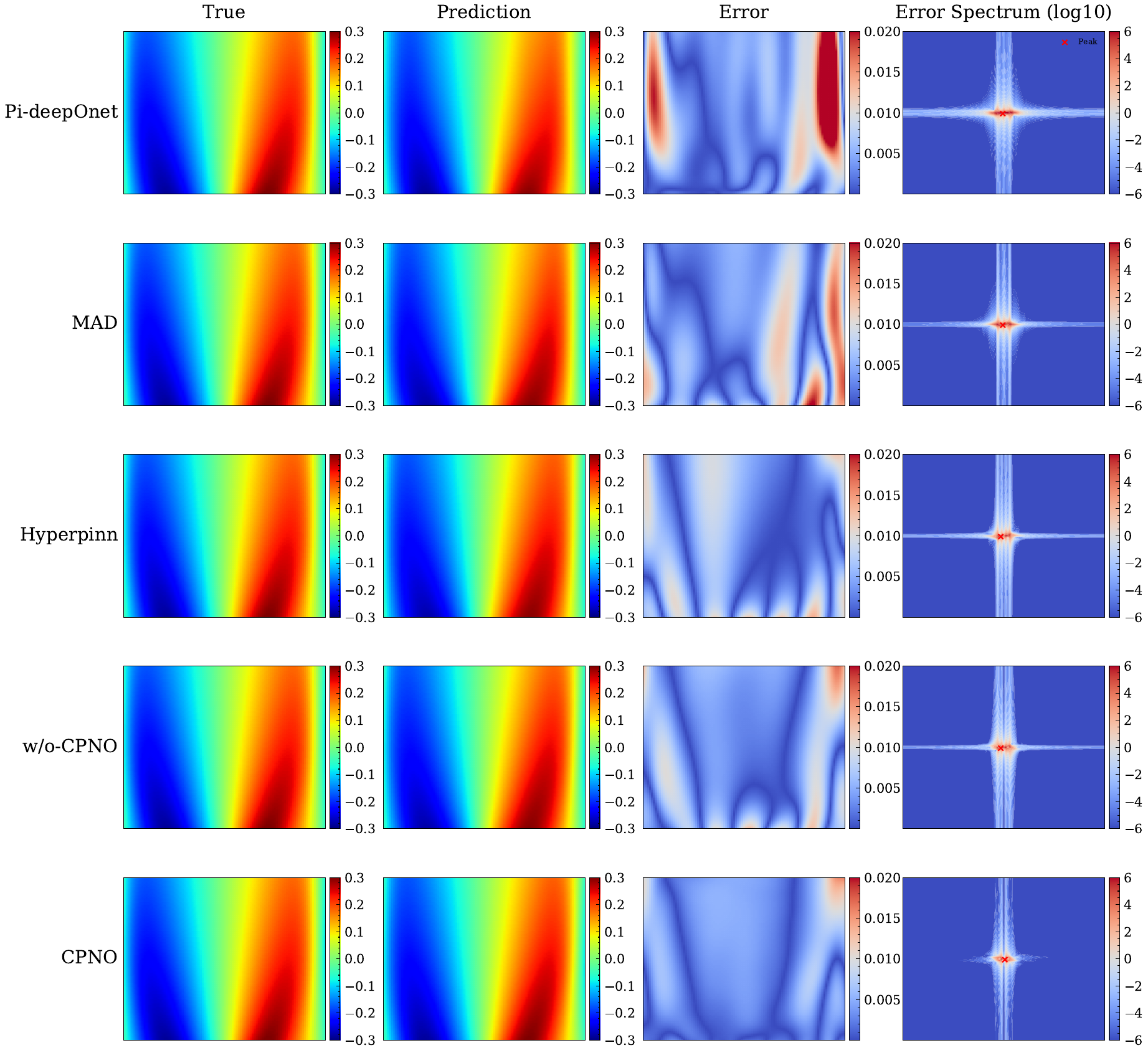} \hfill
    \caption{Visual Comparison of Predicted Solutions, Error Contours, and Error Spectrum Contours for Burgers.}
    \label{fig:visual-burgers}
\end{figure}

Figure~\ref{fig:visual-burgers} offers a comparative visualization, for a selected parameter instance of the Burgers' equation, depicting: (i) the predicted solutions by each method; (ii) contour plots of the point-wise absolute error $| u_{pred}-u_{true} |$ between the predicted and true solutions; and (iii) contour plots illustrating the spatial Fourier spectrum energy distribution of this point-wise error field. Direct observation of the predicted solutions allows for an initial assessment of each method's ability to resolve key features, such as shock waves. The error contour plots further underscore the advantages of CPNO in overall error control; its error contours are generally lighter, indicating lower error magnitudes. Notably, CPNO maintains smaller prediction errors even in regions characterized by high solution gradients, such as near shock fronts, whereas some baseline methods may exhibit significant error concentrations in these critical areas. The error spectrum contour plots provide profound insights into the error distribution across different frequency modes in the spatial domain. These spectra clearly reveal that, compared to the other methods, CPNO not only achieves marked suppression of errors in the high-frequency spatial components—indicative of its superior ability to capture fine details and high-wavenumber solution content—but also generally exhibits lower error levels in the low-frequency regime. This comprehensive spectral error reduction substantiates the capability of CPNO to attain more fidelity approximations of Burgers' equation solutions across the entire frequency spectrum than baselines.

\begin{figure}[htbp]
    \centering    
     \includegraphics[width=1.0\textwidth]{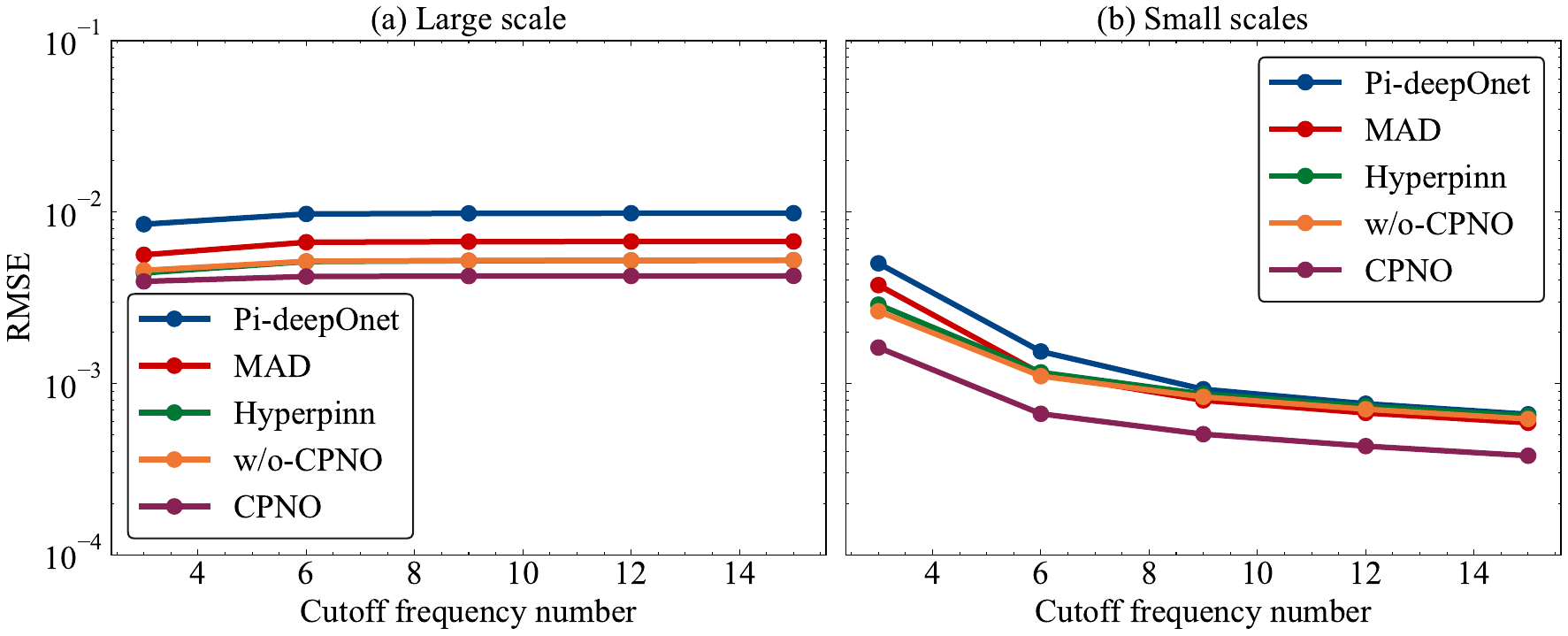} \hfill
    \caption{Multi-scale RMSE Comparison for Burgers. (a) Errors in large-scale components versus cutoff wavenumber; (b) Errors in small-scale components versus cutoff wavenumber}
    \label{fig:scaled-error-burgers}
\end{figure}

To further investigate the capability of different methods in capturing solution structures across various scales for Burgers' equation, we conducted a multi-scale error decomposition of the prediction results. By applying low-pass and high-pass filters in the Fourier domain with varying cutoff wavenumbers, both the true solutions and the predicted solutions from each model were decomposed into large-scale (low-frequency) and small-scale (high-frequency) components. Subsequently, the Root Mean Square Error (RMSE) was computed separately for these two scale components to quantify the prediction accuracy of each model at different scales. Figure~\ref{fig:scaled-error-burgers} presents a comparative analysis of these multi-scale errors for a test case of the Burgers, contrasting CPNO with baselines. Figure~\ref{fig:scaled-error-burgers}a displays the RMSE for the large-scale components, while Figure~\ref{fig:scaled-error-burgers}b shows the RMSE for the small-scale components, both as a function of the cutoff wavenumber. A clear observation from these results is that CPNO consistently exhibits the lowest prediction error across all examined cutoff wavenumbers, for both large-scale and small-scale components. This indicates that CPNO not only accurately captures the global, macroscopic structure of the solution (large-scale features) but also demonstrates superior capability in resolving fine-grained details and high-frequency oscillations (small-scale features). This outcome robustly demonstrates the comprehensive performance advantage and versatility of our proposed method in addressing multi-scale problems.

\subsection{Hyperparameter Sensitivity Analysis}
To comprehensively evaluate the performance of our proposed CPNO and to rigorously validate the efficacy of its core component, we conduct a comparative analysis against its ablation model, w/o-CPNO. Both architectures are evaluated under the previously introduced zero-shot and few-shot training paradigms, yielding four model variants for comparison: CPNO-zero, CPNO-few, w/o-CPNO-zero, and w/o-CPNO-few. This section presents a detailed analysis of these four variants across a range of hyperparameter settings to assess their performance and robustness. We investigate the sensitivity of these models to key hyperparameters, including batch size and learning rate. Batchsize were selected from the set {1000,2000,4000}, and learning rates were chosen from {0.005,0.001,0.0005}, resulting in a total of 9 distinct hyperparameter configurations. For each configuration, all models were independently trained and evaluated on the burgers, with the relative $L_2$ test errors recorded.


\begin{figure}[htbp]
  \centering
  \begin{minipage}[t]{0.48\textwidth}
    \centering
    \includegraphics[width=\linewidth]{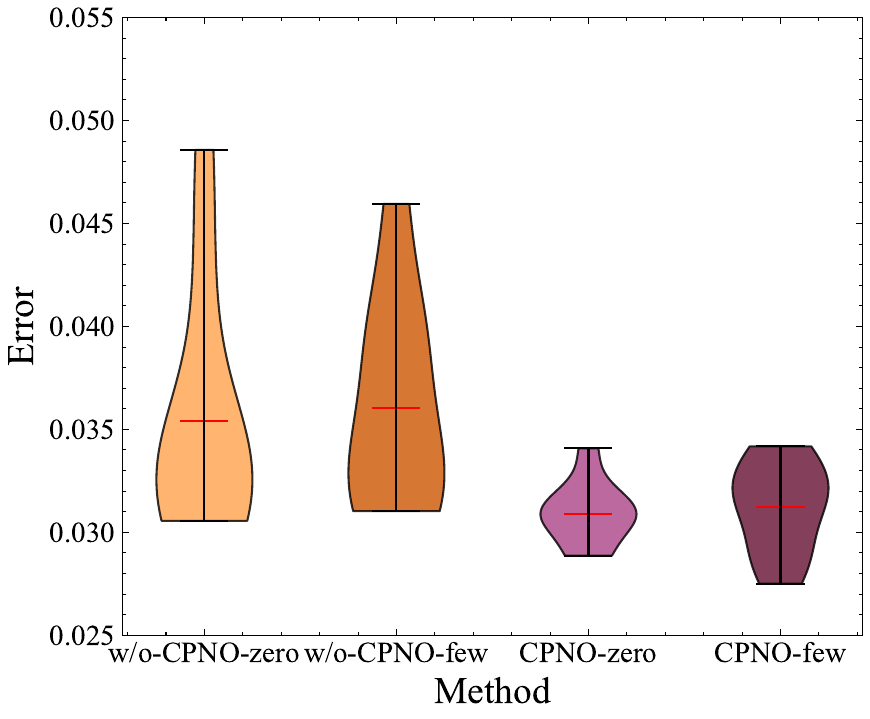}
    \caption{Comparison of Mean Relative $L_2$ Errors and Standard Deviations CPNO and m-PINO across Various Hyperparameter Configurations}
    \label{fig:robustness-error}
  \end{minipage}
  \hfill
  \begin{minipage}[t]{0.48\textwidth}
    \centering
    \includegraphics[width=\linewidth]{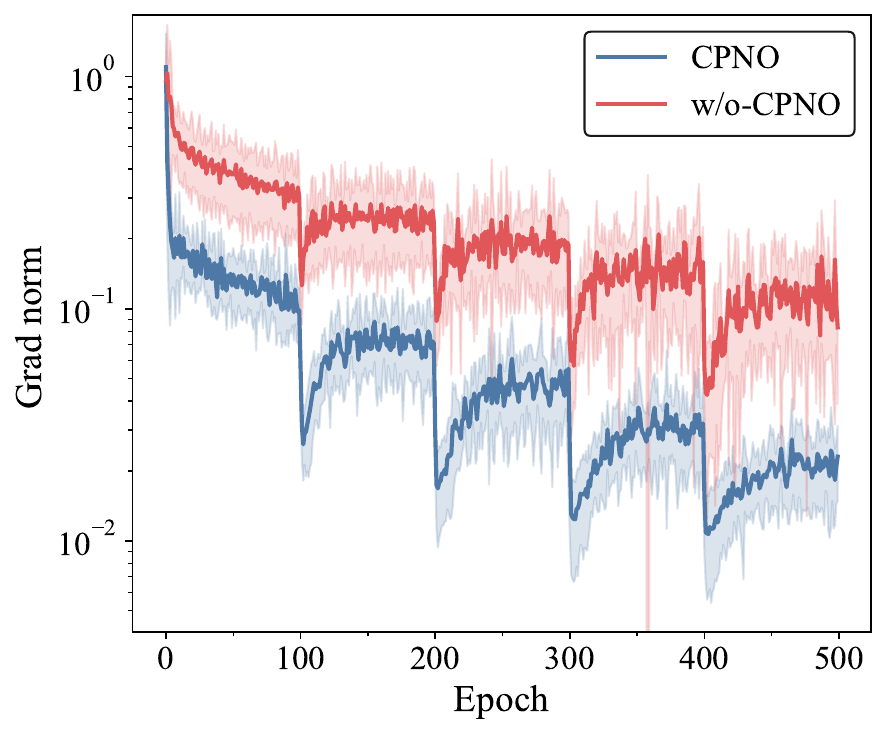}
    \caption{Comparison of Epoch-wise Average $L_2$ Gradient Norms for CPNO and w/o-CPNO during Training across Various Hyperparameter Configurations.}
    \label{fig:hyper-grad}
  \end{minipage}
\end{figure}

Figure~\ref{fig:robustness-error} provides a detailed view of hyperparameter robustness by presenting violin plots of the relative $L_2$ test errors for the four model variants, computed across the nine distinct hyperparameter configurations. The shape of each violin illustrates the full probability density distribution of the errors, with its width corresponding to the density of results at a given error level. This visualization offers a more complete picture of performance consistency than summary statistics alone. A clear observation is that the violins corresponding to both CPNO-zero and CPNO-few are significantly more compact and concentrated at lower error values. This indicates not only a superior median accuracy but also a high degree of performance consistency across the tested hyperparameter range. In contrast, the violins for the w/o-CPNO variants are visibly elongated and wider, signifying a greater sensitivity to hyperparameter selection. The presence of a longer upper tail in these plots suggests that certain hyperparameter choices lead to a substantial degradation in performance, a risk not fully captured by a simple standard deviation metric. These results provide strong evidence that the introduction of Chebyshev spectral encoding contributes to a more stable and robust optimization landscape, significantly mitigating the model's sensitivity to hyperparameter variations.


To further elucidate the impact of Chebyshev encoding on model training dynamics and stability, we compare the evolution of epoch-wise average $L_2$ gradient norms for CPNO and w/o-CPNO across the 9 hyperparameter configurations previously detailed and the average $L_2$ norm of the gradients was recorded at each training epoch. Figure~\ref{fig:hyper-grad} presents the mean curves of these epoch-wise average $L_2$ gradient norms, along with their corresponding standard deviations (indicated by shaded regions), for both CPNO and w/o-CPNO, aggregated over all nine hyperparameter configurations. The abscissa represents the training epoch, while the ordinate depicts the average $L_2$ gradient norm on a logarithmic scale. It is clearly observable from the figure that CPNO not only exhibits a generally lower magnitude of average gradient norms throughout the training process but also demonstrates a significantly smaller standard deviation compared to the w/o-CPNO method. Specifically, the gradient norm curve for w/o-CPNO shows considerable fluctuations and maintains a higher mean and wider variance across different hyperparameter settings, potentially indicating sensitivity in its optimization landscape. In contrast, CPNO's gradient norm curve is substantially smoother, characterized by lower mean values and markedly reduced standard deviation. This suggests that its training process is more stable and less susceptible to variations in hyperparameter choices. Smaller and more stable gradient norms typically correlate with a more amenable loss landscape and a more reliable optimization trajectory, which aligns with the superior accuracy and hyperparameter robustness exhibited by CPNO as reported in Figure~\ref{fig:hyper-grad}.

\subsection{Robustness Analysis: Stability to Input Parameter Noise}

An ideal neural operator must not only yield accurate solutions for precise parameters but also exhibit robustness to small perturbations in its inputs. To systematically evaluate the performance of our proposed CPNO in this regard, we designed a noise robustness experiment using the Burgers' equation as a benchmark. In this experiment, we introduced varying levels of Gaussian white noise to the parameters that define the initial conditions, with standard deviations \(\sigma\) set to \{0.005, 0.01, 0.03\}, and compared these against the noise-free case. We assessed the performance of our CPNO model against its ablation variant, w/o-CPNO, under these noisy input conditions. All models were trained within the purely physics-informed (zero-shot) paradigm.

\begin{figure}[htbp]
    \centering    
     \includegraphics[width=1.0\textwidth]{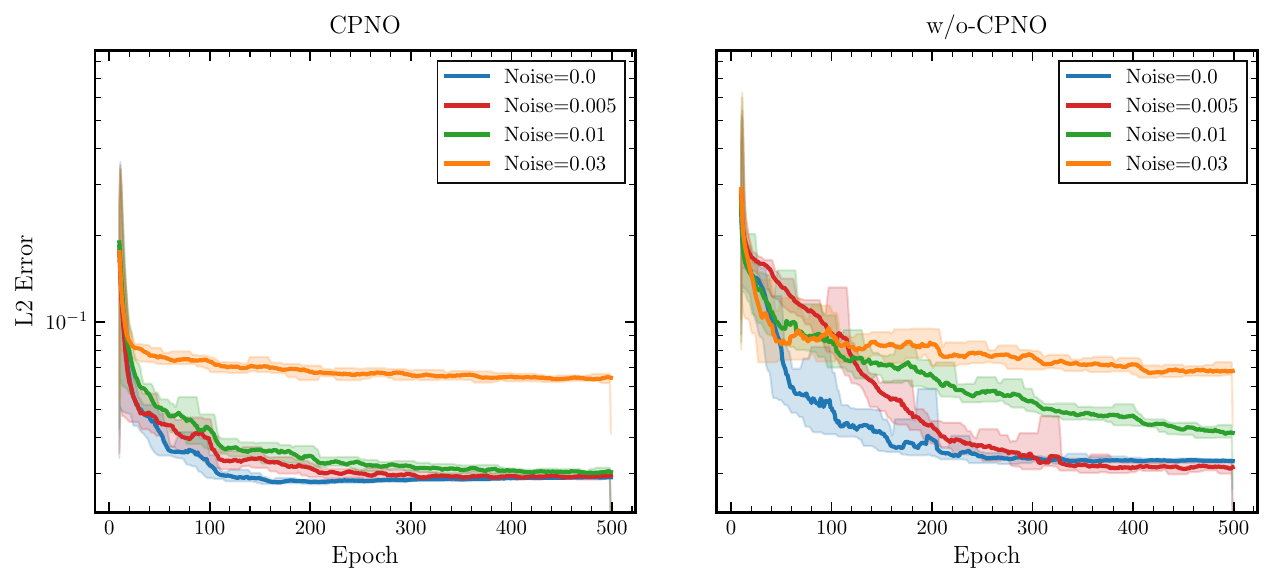} \hfill
    \caption{Training convergence of CPNO and w/o-CPNO for the Burgers' equation under different levels of noise applied to the parameters.}
    \label{fig:noise_convergence}
\end{figure}

Figure~\ref{fig:noise_convergence} illustrates the training convergence process for both models under different noise levels. It is clearly observable that the convergence curves for CPNO at low noise levels are nearly indistinguishable, converging rapidly and stably to a very low error, which indicates that low-level noise has a negligible impact on its training dynamics and final accuracy. While higher noise levels lead to an increased final error, the convergence remains swift. In stark contrast, the convergence curves of w/o-CPNO demonstrate a significant sensitivity to all noise levels. As the noise level increases, its convergence process becomes more oscillatory, and the final error is substantially higher, indicating that its training process is severely perturbed by the parameter disturbances.

\begin{figure}[htbp]
    \centering    
     \includegraphics[width=1.0\textwidth]{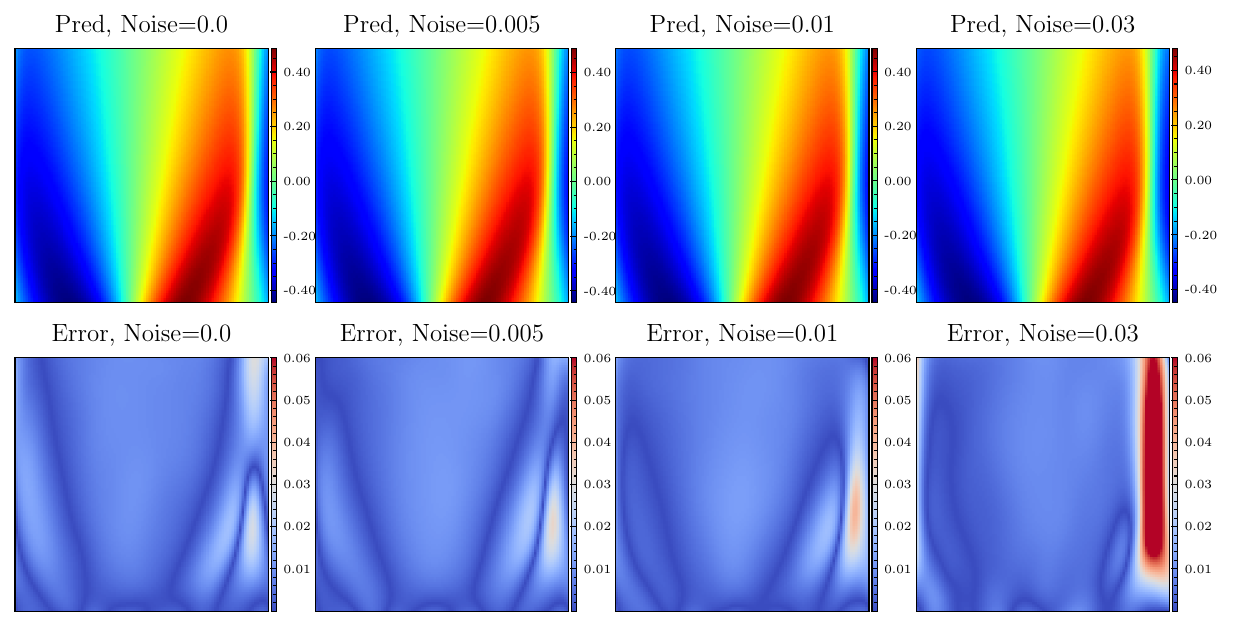} \hfill
    \caption{ Predicted solutions (top row) and pointwise error fields (bottom row) by CPNO under varying levels of input noise.}
    \label{fig:noise_viz_CPNO}
\end{figure}

\begin{figure}[htbp]
    \centering    
     \includegraphics[width=1.0\textwidth]{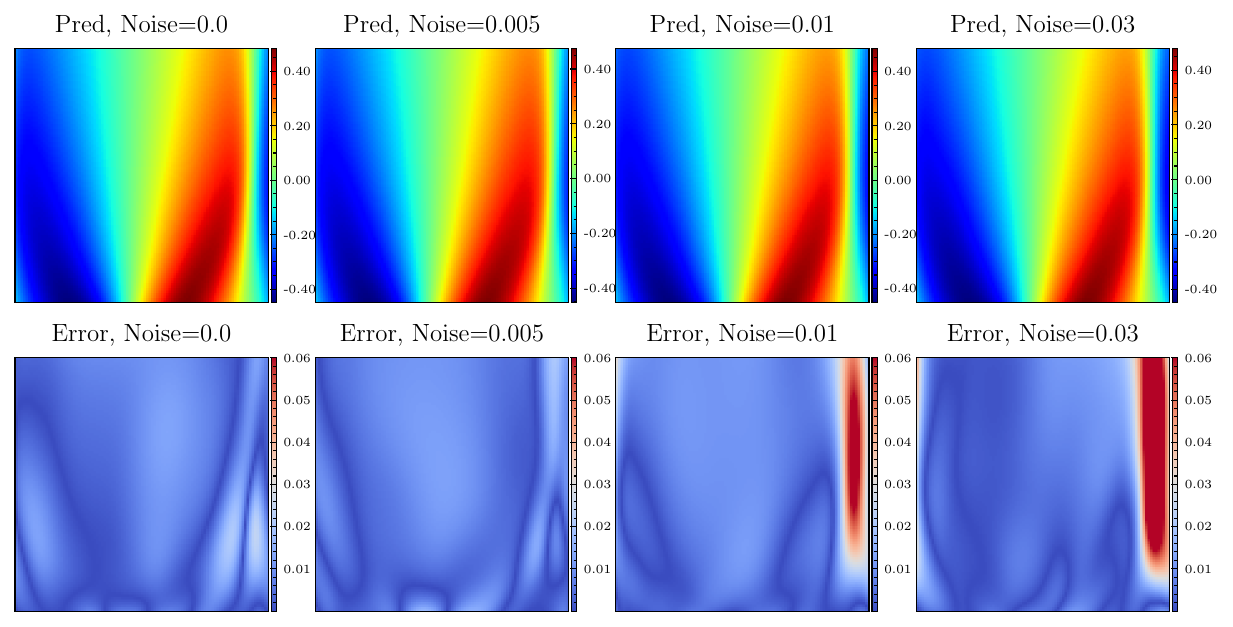} \hfill
    \caption{ Predicted solutions (top row) and pointwise error fields (bottom row) by w/o-CPNO under varying levels of input noise.}
    \label{fig:noise_viz_woCPNO}
\end{figure}

To provide a more intuitive understanding of this performance disparity, Figures~\ref{fig:noise_viz_CPNO} and \ref{fig:noise_viz_woCPNO} present a qualitative comparison of the predicted solutions and pointwise error fields for a test case under varying noise levels. As shown in Figure~\ref{fig:noise_viz_CPNO}, the predictions of CPNO are visually almost identical to the noise-free prediction, and its error field remains at an extremely low magnitude across all noise levels. This demonstrates that CPNO has successfully learned a stable operator mapping. Figure~\ref{fig:noise_viz_woCPNO}, however, reveals the fragility of the w/o-CPNO model. As the noise level increases, the quality of its predicted solution degrades sharply, with significant, non-physical oscillations appearing, particularly near the shock front. The corresponding error fields exhibit large regions of high error, confirming that the learned operator mapping is unstable. These results demonstrate the robustness of our operator framework to low-level noise, which is primarily attributed to the numerically stable feature space provided by the Chebyshev polynomials. This enables the operator to learn a smoother parameter-to-solution manifold that is insensitive to input perturbations.

\subsection{Ablation Study}
\subsubsection{Polynomial Layer Depth and Feature Representation }
\begin{figure}[htbp]
    \centering    
     \includegraphics[width=1.0\textwidth]{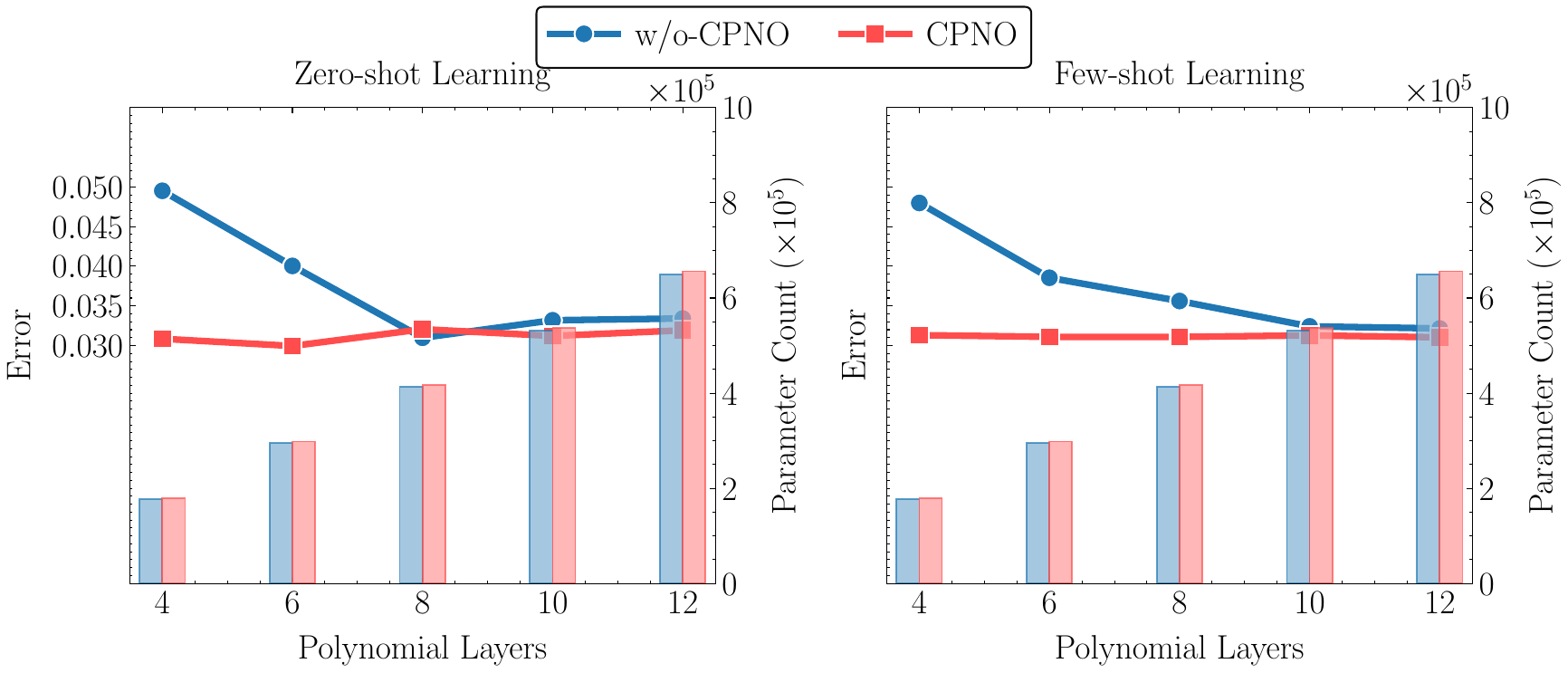} \hfill
    \caption{ Comparison of final test errors for Different polynomial layer Depth on case burgers equation}
    \label{fig:inr-layers}
\end{figure}

To investigate the influence of the synthesis network's depth on model performance and to further elucidate the synergistic relationship between the Chebyshev spectral encoding and the network architecture, we conduct an ablation study. We evaluate the performance of four model variants (w/o-CPNO-zero, w/o-CPNO-few, CPNO-zero, and CPNO-few) across a range of network depths, with the number of layers set to {4, 6, 8, 10, 12}. All experiments are performed on the Burgers' equation.

Figure~\ref{fig:inr-layers} illustrates the trends of model accuracy and network parameter count as a function of depth, presented in two subplots for the zero-shot and few-shot paradigms, respectively. In each subplot, the horizontal axis represents the number of layers in the synthesis network, while the left and right vertical axes denote the final relative $L_2$ test error and the total number of network parameters, respectively. Several key phenomena are observed. For the w/o-CPNO model, which processes raw coordinates directly, a relatively high error is noted for shallow networks (e.g., 4 layers), indicating that a shallow network struggles to construct sufficiently complex features from the raw inputs. While its error gradually decreases as the network deepens, this improvement comes at the cost of a linear increase in the number of parameters. In stark contrast, the CPNO model, which incorporates Chebyshev encoding, exhibits a distinctly different behavior. CPNO achieves a very low error level even with a shallow network (e.g., 4 or 6 layers). Further increases in network depth yield only marginal improvements in accuracy, with its performance curve remaining flat and at a low error level across a wide range of depths.

This comparison provides compelling evidence for a core advantage of the Chebyshev spectral encoding: a well-designed, information-rich input feature space significantly reduces the model's dependency on network depth. As shown in Figure X(a), a 6-layer CPNO-zero model achieves an accuracy far superior to that of a 12-layer w/o-CPNO-zero model, while possessing only half the number of parameters. This clearly demonstrates that by performing an efficient spectral feature transformation at the input stage, we can achieve or even surpass the performance of a much deeper and more complex network with a more compact and efficient architecture. Therefore, Chebyshev encoding not only directly improves the model's final accuracy but, more importantly, enhances the quality of feature representation, making it possible to construct lighter and more parameter-efficient neural operators.

\subsubsection{Chebyshev Orders for Solution Mapping}

\begin{figure}[htbp]
    \centering    
     \includegraphics[width=0.6\textwidth]{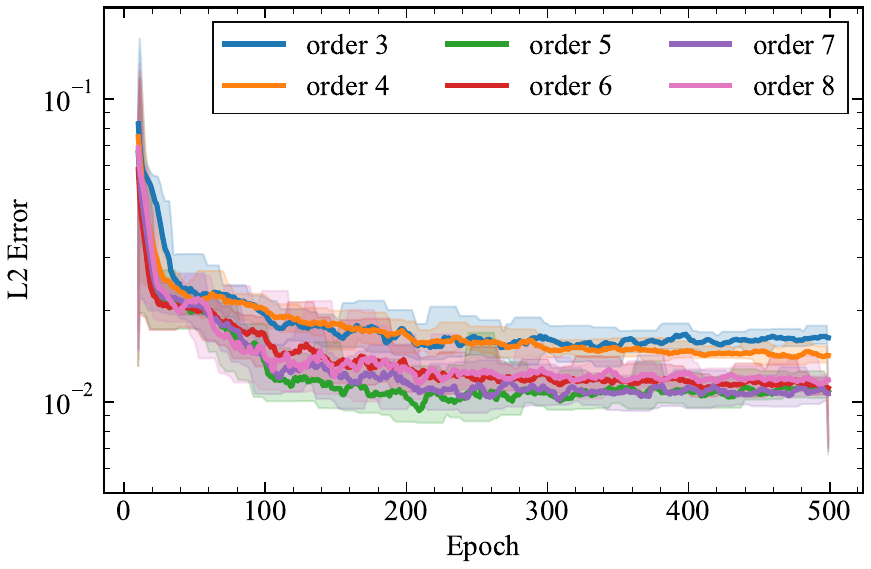} \hfill
    \caption{ Comparison of final test errors for chebyshev polynomial orders on case Allen-cahn equation}
    \label{fig:cheby-orders}
\end{figure}

A key hyperparameter in our proposed method is the maximum degree of the Chebyshev polynomials used in the spectral encoding module. To investigate the model's performance sensitivity to this parameter, we conducted an ablation study on the Burgers equation. For this experiment, the depth of the synthesis network was fixed at 8 layers, and we evaluated the performance of the CPNO-zero model while systematically varying the Chebyshev order $P \in {3,4,5,6,7,8}$. Figure~\ref{fig:cheby-orders} illustrates the final relative $L_2$ test error as a function of the encoding order. A clear trend is observed where the error decreases markedly as P is increased from 3 to 5. This indicates that lower-order spectral features are insufficient to fully capture the complexity of the Burgers' solution manifold, and that incorporating higher-order spectral information is essential for achieving high accuracy. Critically, for $P>5$, the error curve becomes nearly flat, demonstrating that the model's performance converges.

This result reveals the existence of an effective convergence order for the Chebyshev encoding for this problem. It demonstrates the robustness of our method, as its performance is not infinitely sensitive to the choice of P and it stabilizes beyond a certain sufficient order. A finite and moderate number order of chebyshev polynomial  are sufficient to accurately represent the complex solution manifold, representing a balance between expressive power and computational cost. 

\subsection{Extension to Complex Geometries: Transonic Airfoil Flow}

While preceding experiments have demonstrated the robust zero-shot solving capabilities of CPNO for benchmark parameterized PDEs, these problems are predominantly confined to regular domains (e.g., rectangular or periodic boundaries). To further validate CPNO's efficacy in irregular engineering geometries, this section employs a data-driven approach to evaluate its performance on transonic airfoil flow—a paradigmatic challenge in aerospace engineering characterized by intricate boundaries and multiscale phenomena, such as shock waves and turbulent boundary layers. High-fidelity CFD simulation data are utilized to train CPNO, with assessments focusing on accuracy and generalization in complex geometries.
Transonic airfoil flow is governed by the Reynolds-averaged Navier-Stokes (RANS) equations, subject to far-field freestream and no-slip conditions on the airfoil surface. The dataset comprises approximately 6400 instances, where each instance's input parameters include airfoil geometric descriptors and angle of attack, while the output flow fields encompass pressure and velocity distributions. CPNO is configured with a polynomial order of $P=10$ and Chebyshev encoding degree of 16. Training follows a data-driven paradigm, minimizing the mean squared error loss $\mathcal{L} = \frac{1}{N} \sum_{i=1}^N \| u_\phi(\mathbf{x}_i; \theta_i) - u_{\text{ref}}(\mathbf{x}_i; \theta_i) \|_2^2$, with evaluation based on the L1 absolute error metric.

\begin{table}[htbp]
\centering
\caption{L1 Absolute Errors of CPNO for Transonic Airfoil Flow Prediction}
\label{tab:error_airfoil}
\begin{tabular}{lcccccc}
\toprule
\multirow{2}{*}{L1 Error} & \multicolumn{2}{c}{u} & \multicolumn{2}{c}{v} & \multicolumn{2}{c}{p} \\
\cmidrule(lr){2-3} \cmidrule(lr){4-5} \cmidrule(lr){6-7}
& mean & max & mean & max & mean & max \\
\midrule
Training Set & 0.00175 & 0.07744 & 0.00094 & 0.08072 & 0.00126 & 0.05493 \\
Test Set & 0.00216 & 0.08727 & 0.00103 & 0.08209 & 0.00121 & 0.05463 \\
\bottomrule
\end{tabular}
\end{table}

\begin{figure}[htbp]
    \centering    
     \includegraphics[width=0.9\textwidth]{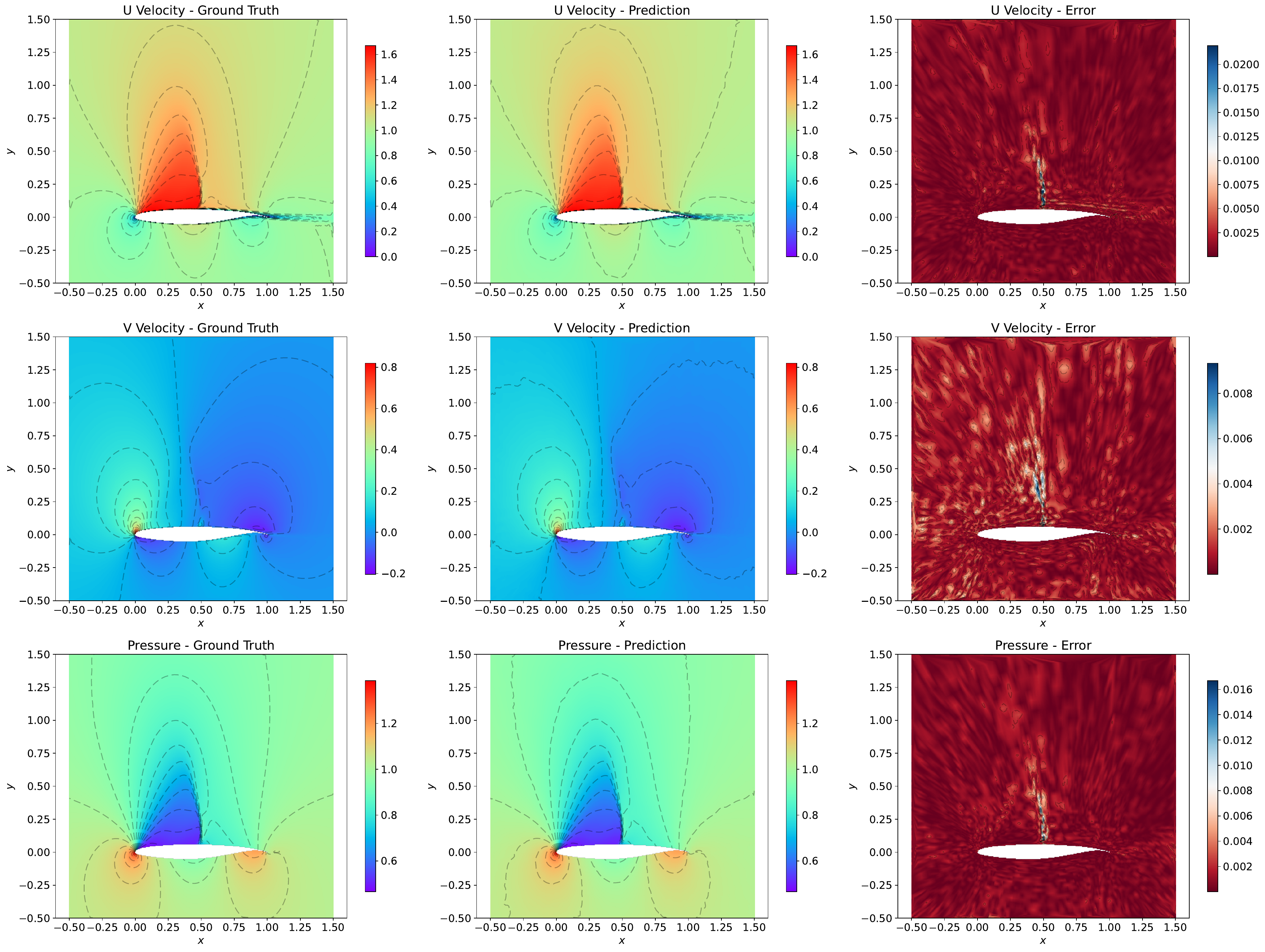} \hfill
    \caption{Comparison of Predicted and Ground-Truth Global Distributions for Velocity Components (u, v) and Pressure (p) Fields}
    \label{fig:airfoils}
\end{figure}

Evaluation results for the training and test sets are presented in Table~\ref{tab:error_airfoil}, revealing mean errors consistently on the order of $10^{-3}$, with maximum errors constrained to the order of $10^{-1}$. The close alignment between training and test errors underscores the algorithm's strong generalization to variations in geometry and inflow conditions. Figure~\ref{fig:airfoils} provides visualizations of model predictions on test cases, demonstrating effective capture of shock structures and highlighting CPNO's high-fidelity forecasting. These findings indicate that CPNO can be effectively extended to complex engineering geometries, offering valuable insights for aircraft design optimization.

\section{Conclusion}
This work presents a novel Physics informed Chebyshev Polynomial Neural Operator (CPNO) aimed at solving parametric partial differential equations (PDEs) accurately and robustly. We construct a stacked operator framework wherein the nonlinearity is represented by a polynomial expansion, thus obviating the need for predefined nonlinear components. We employ Chebyshev polynomials as the foundational basis within each layer to simultaneously enhance the representational capacity and training stability of the network. Through systematic numerical experimentation on benchmark parameterized PDE problems, the performance of CPNO was rigorously evaluated. Results demonstrate that CPNO not only achieves significant advantages in final solution accuracy but also exhibits faster convergence rates compared to representative baseline methods, including PI-DeepONet, MAD, and HyperPINNs. Multi-scale error analysis and error spectrum analysis further substantiate CPNO's exceptional capability in accurately capturing both macroscopic solution structures and fine-grained, high-frequency details. Moreover, assessments of hyperparameter sensitivity and training process gradient norms indicate that CPNO possesses enhanced robustness and training stability.

Notwithstanding the superior performance of CPNO across multiple facets, the current framework exhibits certain limitations. When processing variable parameters that are unstructured, it presently relies on interpolation methods, as its coordinate representation and network architecture are better suited to regular or simply parameterizable domains. A significant challenge for future research lies in effectively extending CPNO, along with its associated encoding and modulation strategies, to arbitrary complex geometries. Several promising avenues for future investigation emerge. Firstly, the integration of data-driven strategies, by leveraging extensive high-fidelity data, could guide the network in learning behaviors on complex geometric domains or serve to calibrate and enhance the predictive accuracy of the underlying physical model. Secondly, for challenging problems involving higher-dimensional parameter spaces and spatiotemporal domains, further optimization of the CPNO architecture and training algorithms is warranted to improve scalability and computational efficiency, thereby better mitigating the curse of dimensionality. Finally, extending the applicability of CPNO to solve dynamical systems exhibiting chaotic behavior constitutes another compelling research direction. Such problems impose stringent demands on long-term predictive stability and the accurate capture of sensitivity to initial conditions and parameters, potentially necessitating the incorporation of advanced time-series modeling techniques or uncertainty quantification methods.

\bibliography{main.bib}

\appendix{Appendix}
\section{Parametric PDE Formulations}
\label{app:equations}
This appendix provides detailed formulations of the parametric partial differential equations (PDEs) evaluated in Section 4.1, including the Burgers equation, the Allen-Cahn equation, and a nonlinear diffusion-reaction system. These equations model diverse physical phenomena, and their parameter variations were generated using a Gaussian Random Field to test the performance of the proposed CPNO framework.

\textbf{Burgers Equation}

Burgers equation captures the dynamics of incompressible, viscous fluid without external forces, which is instrumental in understanding the development of turbulence and the formation of shock waves. We consider the following Burgers equation with various initial conditions and periodic boundary condition:
\begin{equation}
\begin{cases}
\frac{d s}{d t}+s \frac{d s}{d x}-\nu \frac{d^2 s}{d x^2}=0,\quad (x, t) \in[0,1]^2  \\
s(x, 0)=u(x),  \\
s(0, t)=s(1, t), \quad \frac{d s}{d x}(0, t)=\frac{d s}{d x}(1, t)
\end{cases}
\end{equation}
Our goal is to map the initial conditions to the solutions of the equation. The initial conditions $u(x)$ are generated randomly for 900 samples and 10 samples for test.

\textbf{Diffusion-reaction system}

The nonlinear diffusion-reaction system capture how reactants and products evolve over time and space due to diffusion and reaction processes. We consider the following nonlinear diffussion-reaction systems with zero initial and boundary conditions:
\begin{equation}
\begin{cases}
\frac{\partial s}{\partial t}=D \frac{\partial^2 s}{\partial x^2}+k s^2+u(x), \quad(x, t) \in(0,1]^2 \\
s(0,t) = 0, s(x,0)=0
\end{cases}
\end{equation}
where $D = 0.01$ is the diffusion coefficient and $k = 0.01$ is the reaction rate. Our goal is to map the source term $u(x)$ to the PDE solutions $s(x)$. We generate 1000 cases of source term $u(x)$ for training and 10 cases for test.

\textbf{Allen-Cahn Equation}

The Allen-Cahn equation is a fundamental reaction-diffusion equation widely used to model phase separation phenomena in materials science, such as the process of coarsening in binary alloys, and it describes the temporal evolution of a scalar order parameter. We consider the following Allen-Cahn equation with various initial conditions and periodic boundary conditions:
\begin{equation}
\begin{cases}
\partial_t u = d_1 \Delta u + d_2 u(1 - u^2), \\
u(0, \boldsymbol{x}) = f_0(\boldsymbol{x}),
\end{cases}
\end{equation}
Here, $\epsilon$ is a parameter controlling the interface width. Our objective is to learn the mapping from diverse initial conditions $u(x)$ to the corresponding solutions $s(x,t)$. The initial conditions $u(x)$ are randomly generated, with 900 samples for training and 10 samples for test.

\textbf{Navier-Stokes Equations}
The vorticity formulation of Navier-Stokes equations describes the spatiotemporal evolution of the scalar vorticity field and the associated divergence-free velocity in viscous flows. We consider the following two-dimensional incompressible Navier-Stokes equations in vorticity-velocity form with periodic boundary conditions:

\begin{equation}
\begin{cases}
\partial_t w(\boldsymbol{x}, t)+u(\boldsymbol{x}, t) \cdot \nabla w(\boldsymbol{x}, t)=\nu \Delta w(\boldsymbol{x}, t)+f(\boldsymbol{x}), \\
w(\boldsymbol{x}, 0)=w_0(\boldsymbol{x})
\end{cases}
\end{equation}
with source $f(\boldsymbol{x})= 0.1(\sin (2 \pi(x+y))+\cos (2 \pi(x+y)))$ and $\nu = 0.1$. Our goal is to learn the map $w(x, t)$ to $w(x, t)$ with $\Delta t=0.1s$. We generate 500 cases like FNO\cite{li2020fourier} with resolution $32*32$.

\section{Proof of the Truncation Error of CPNO}
The truncation error associated with the Chebyshev basis and the coefficient estimation error introduced by the neural network:

\begin{equation}
E_{tru} = \left\| u(x; \theta) - \sum_{k=0}^N c_k^*(\theta) T_k(x) \right\|_{L^\infty}
\end{equation}
where \(e_k^*\) are the true Chebyshev coefficients obtained via the weighted inner product:
\begin{equation}
c_k^* = \frac{\langle u(x; \theta), T_k \rangle}{\langle T_k, T_k \rangle}, \quad \langle f, g \rangle = \int_{-1}^{1} \frac{f(x) g(x)}{\sqrt{1 - x^2}} \, dx
\end{equation}
and the orthogonality of the Chebyshev polynomials satisfies \(\langle T_m, T_n \rangle = \frac{\pi}{2} \delta_{mn}\) (with \(\langle T_0, T_0 \rangle = \pi\) when \(m, n \neq 0\)).

The truncation error assesses the intrinsic capability of the Chebyshev basis to approximate \(u(x; \theta)\) using a finite \(N\)-term series. The true coefficients \(e_k^*\) are computed via orthogonal projection:

\begin{equation}
c_k^* = \frac{\int_{-1}^{1} u(x; \theta) T_k(x) w(x) \, dx}{\int_{-1}^{1} T_k(x)^2 w(x) \, dx}
\end{equation}
where the weight function \(w(x) = 1/\sqrt{1 - x^2}\) ensures orthogonality, and the denominator \(\langle T_k, T_k \rangle = \pi/2\) for \(k \geq 1\) (with \(\langle T_0, T_0 \rangle = \pi\)). Consequently, \(u(x; \theta)\) admits an infinite series representation \(u(x; \theta) = \sum_{k=0}^\infty e_k^* T_k(x)\), and the truncation error corresponds to the tail \(\sum_{k=N+1}^\infty e_k^* T_k(x)\).

Given that \(u(x; \theta)\) is a smooth function, we hypothesize its analyticity within a Bernstein ellipse with foci at \(\pm 1\) and semi-major axis, which allows \(u(x; \theta)\) to be extended as a holomorphic function in this ellipse, implying that the Chebyshev coefficients \(e_k^*\) decay exponentially with \(k\), i.e., \(|c_k^*| \leq C \rho^{-k}\) for some \(C > 0\) and \(\rho > 1\), where \(\rho\) depends on the radius of analyticity.
The truncation error is bounded by evaluating the \(L^\infty\) norm of the tail series. Leveraging the uniform boundedness of Chebyshev polynomials (\(|T_k(x)| \leq 1\) for \(x \in [-1, 1]\)), we obtain:

\begin{equation}
\left| \sum_{k=N+1}^\infty c_k^* T_k(x) \right| \leq \sum_{k=N+1}^\infty |c_k^*| \cdot |T_k(x)| \leq \sum_{k=N+1}^\infty C \rho^{-k}.
\end{equation}

This is a geometric series, and since \(\rho > 1\), it converges. The sum is:

\begin{equation}
\sum_{k=N+1}^\infty \rho^{-k} = \rho^{-(N+1)} \cdot \frac{1}{1 - \rho^{-1}} = \frac{\rho^{-(N+1)}}{\rho - 1}.
\end{equation}

Substituting the coefficient bound, we get:

\begin{equation}
\sum_{k=N+1}^\infty C \rho^{-k} = C \cdot \frac{\rho^{-(N+1)}}{\rho - 1}.
\end{equation}

Noting that \(\rho^{-(N+1)} = \rho^{-N} \cdot \rho^{-1}\), and adjusting the constant \(C\) to account for the initial shift and the supremum of \(u\) over the ellipse, the bound simplifies to:

\begin{equation}
\left\| u(x; \theta) - \sum_{k=0}^N c_k^* T_k(x) \right\|_{L^\infty} \leq \sup_{x \in [-1, 1]} \left| \sum_{k=N+1}^\infty c_k^* T_k(x) \right| \leq \frac{C \rho^{-N}}{\rho - 1},
\end{equation}
where $C$ is rescaled to reflect the maximum value of $u$ in the analytic region. This result confirms the spectral convergence of the Chebyshev basis, distinguishing it from the algebraic convergence of traditional polynomial bases.

\section{Analysis of Gram Matrices for Nonlinear Differential Operators}

To empirically validate the preceding theoretical analysis and to reveal the disparity in numerical stability between the two bases in a more practical scenario, we designed a key numerical experiment. We employ a damped Laplacian operator as the test case:
\begin{equation}
    \mathcal{L}[u] = \alpha \frac{d^2u}{dx^2} + u, \quad \text{with }   \alpha=0.01.
\end{equation}
The choice of this operator is highly representative. Unlike the simplest identity operator, the inclusion of a second-order derivative term allows the operator to act on different scales, thus providing a more comprehensive test of the basis's stability under differentiation. Concurrently, by selecting a small $\alpha$, we ensure that the impact of the differential term is sufficient to distinguish the performance of the two bases without being so large as to cause immediate failure from numerical overflow for the monomial basis. We construct the Gram matrix $\mathbf{G}$ for both the monomial basis $\{\psi_k(x) = x^k\}$ and the Chebyshev basis $\{\psi_k(x) = T_k(x)\}$, with elements defined by the integral $G_{jk} = \langle \mathcal{L}[\psi_j], \mathcal{L}[\psi_k] \rangle_{L^2}$. High-precision Gauss-Legendre quadrature is used for the numerical integration, and we analyze the condition number $\kappa(\mathbf{G})$ as the maximum polynomial degree $N$ varies from 2 to 14.

\begin{figure}[h!]
 \centering
 \includegraphics[width=0.6\textwidth]{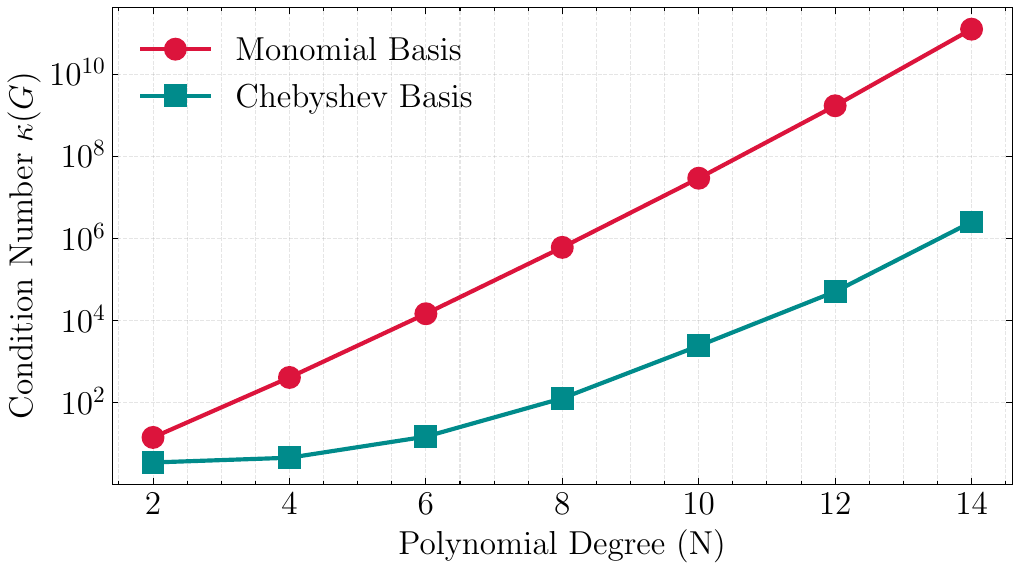}
 \caption{A comparison of the condition number $k(G)$ as a function of the polynomial order $N$.}
 \label{fig:cond-comparison}
 \end{figure}

Figure ~\ref{fig:cond-comparison} clearly reveal a significant chasm in the numerical stability of the two bases, the condition number of the Gram matrix for the monomial basis, $\kappa(\mathbf{G}_M)$, exhibits an explosive exponential growth with respect to the degree $N$. At $N=14$, its condition number reaches a staggering $10^{11}$, rendering any numerical computation based on this matrix utterly unreliable due to the catastrophic amplification of round-off errors. In stark contrast, the condition number for the Chebyshev basis, $\kappa(\mathbf{G}_T)$, while also increasing, grows at a much slower rate, following a low-order polynomial trend.

It is noteworthy that, compared to the identity operator case, the condition number for the Chebyshev basis does increase. This is because the introduction of the differential operator $\alpha u''$ perturbs the near-orthogonal structure of the basis under the standard $L^2$ inner product. However, the most critical observation lies in the order-of-magnitude difference in the growth rates. The intrinsic stability of the Chebyshev basis ensures that the degradation in conditioning caused by this differential perturbation is controlled and polynomial. Conversely, the monomial basis is already inherently ill-conditioned, and the differential operator merely exacerbates its intrinsic exponential instability. At $N=14$, the ratio between the two condition numbers is on the order of $10^5$, clearly demonstrating the absolute superiority of the Chebyshev basis. 
This numerical experiment provides compelling empirical evidence for our core theory: due to its superior intrinsic mathematical properties, the Chebyshev basis leads to a significantly more stable optimization problem when constructing learning frameworks involving differential operators. This solidifies its role as the foundational choice for high-order, high-precision operator learning.


\end{document}